\documentclass[12pt]{iopart}

\usepackage{iopams}
\usepackage{graphicx,amssymb}
\usepackage{color}
\newtheorem{lemma}{Lemma}
\newtheorem{proposition}{Proposition}

\newcommand{\be}{\begin{equation}}
\newcommand{\ee}{\end{equation}}
\def\dif{\mathop{\rm d\!}\nolimits}

\begin{document}

\title[Thermodynamic class II Szekeres-Szafron regular models]
{Thermodynamic class II Szekeres-Szafron solutions. Regular models}

\author{Bartolom\'e Coll$^{1}$, Joan Josep Ferrando$^{1,2}$ and\\ Juan Antonio S\'aez$^3$}

\address{$^1$\ Departament d'Astronomia i Astrof\'{\i}sica, Universitat
de Val\`encia, E-46100 Burjassot, Val\`encia, Spain}

\address{$^2$\ Observatori Astron\`omic, Universitat
de Val\`encia, E-46980 Paterna, Val\`encia, Spain}

\address{$^3$\ Departament de Matem\`atiques per a l'Economia i l'Empresa,
Universitat de Val\`encia, E-46022 Val\`encia, Spain}

\ead{joan.ferrando@uv.es; juan.a.saez@uv.es}

\begin{abstract}
In a recent paper (Coll {\em et al} 2019  {\it Class. Quantum Grav.} {\bf 36} 175004) we have studied a family of Szekeres-Szafron solutions of class II in local thermal equilibrium (singular models). In this paper we deal with a similar study for all other class II Szekeres-Szafron solutions without symmetries. These models in local thermal equilibrium (regular models) are analyzed and their associated thermodynamic schemes are obtained. 
In particular, we focus on the subfamily of solutions which are compatible with the generic ideal gas equation of state ($p = \tilde{k} n \Theta$), and we analyze in depth two notable interpretations that follow on from the choice of two specific thermodynamic schemes: firstly, as a generic ideal gas in local thermal equilibrium and, secondly, as a model having the homogeneous temperature of the FLRW limit. The models above are shown to fulfill the general necessary macroscopic requirements for physical reality (positivity of matter density, internal energy and temperature, energy conditions and compressibility conditions) in wide domains of the spacetime. 
\end{abstract}
%

%


\section{Introduction}
\label{sec-intro}

An important task in Relativity is to study the physical interpretation of the formal perfect fluid solutions to the Einstein field equations. At present, wide families of
such solutions are known but lack a specific physical meaning. Many of them have been obtained by imposing geometric constraints that simplify the integration of the field equations: static or stationary solutions, or invariance under other isometry groups such as spherically symmetric solutions; or restrictions on the curvature tensor, as in the case of algebraically special solutions. Kinematic constraints on fluid flow have also been imposed a priori in looking for new solutions: geodesic, irrotational, shear-free or non-expanding solutions have been considered. In any case, very few solutions have been interpreted as physically realistic fluids. 

Energy conditions \cite{Plebanski} are necessary constraints for physical reality imposed on the perfect fluid energy tensor $T$. But complementary physical requirements must also be imposed if we look for solutions that describe thermodynamic perfect fluids in local thermal equilibrium (l.t.e.). We must add to the {\em hydrodynamic} quantities that appear in the energy tensor $T = (\rho+ p) u \otimes u + p \, g$, a set of {\em thermodynamic} quantities constrained by the usual thermodynamic laws \cite{Eckart}. Furthermore, to obtain a coherent theory of shock waves, one must impose the relativistic compressibility conditions  \cite{Israel, Lichnero-1}. Our macroscopic hydrodynamic approach to local thermal equilibrium \cite{CFS-LTE} and to relativistic compressibility conditions \cite{CFS-CC} provides a tool to impose these physical requirements.

The Szekeres-Szafron solutions are known as
significant inhomogeneous cosmological models \cite{Krasinski,
Krasinski-Plebanski, EMM, Krasinski-et-all}. These metrics were obtained by Szekeres \cite{Szekeres} for dust solutions and generalized by Szafron \cite{Szafron} for a non-vanishing pressure. The physical and geometric properties of the Szekeres-Szafron (SS) models with constant pressure have been widely analyzed in the literature \cite{Krasinski, Krasinski-Plebanski} (see also the recent papers \cite{hellaby, G-hellaby} and references therein). An invariant approach to these solutions can be found in several papers \cite{Wainwright, Szafron-Collins, Barnes-Row, FS-SS}. 

The study of the physical interpretation of the full set of Szekeres-Szafron metrics is still an open problem. Although some authors have remarked on the difficulties in associating a realistic equation of state to these models \cite{Krasinski-et-all, Lima-Tiomno-a}, there are several results that shed light on this question. Thus, a subfamily of class II SS metrics that evolve to a Friedmann-Lema\^{\i}tre-Robertson-Walker (FLRW) era have been proposed as two-fluid cosmologies \cite{Lima-Tiomno-a}. In a subsequent paper \cite{Lima-Tiomno-b} a subset of the parabolic solutions has been interpreted as one-component fluids in local thermal equilibrium. A wider family with a similar thermodynamic scheme was considered in \cite{QS-1995}. On the other hand, a thermodynamic Szekeres-Szafron solution of class I admits, necessarily, a three dimensional group of isometries on two dimensional orbits \cite{KQS} (see also the recent paper \cite{FS-SS}). Nevertheless, there exist thermodynamic Szekeres-Szafron solutions of class II without symmetries \cite{KQS}.

In \cite{PSS} we have presented a first contribution to the in-depth study of the class II SS spacetimes describing the evolution of a thermodynamic perfect fluid in local thermal equilibrium. We have shown that three families arise naturally: the {\em singular models} (parabolic models with a linear dipole term), the {\em regular models} and the metrics admitting a three-dimensional isometry group $G_3$ on space-like two-dimensional orbits $S_2$. Moreover, we have also accurately analyzed the singular models. In this paper we undertake a similar study for the regular models. 

In section \ref{sec-physical} we introduce the general necessary macroscopic physical requirements that we impose on the solutions, we summarize our results \cite{CFS-LTE, CFS-CC} for the hydrodynamic characterization of the local thermal equilibrium and of the compressibility conditions, and we present a procedure to analyze the general constraints on the physical reality of the known relevant families of perfect fluid solutions. This procedure is based on the important fact, shown in \cite{CFS-LTE}, that a thermodynamic perfect fluid evolving in local thermal equilibrium may be decoupled in a deterministic and independent purely hydrodynamic flow and a set of appropriate thermodynamic quantities associated to it.

In section \ref{sec-classII-termo} we write the canonical form of the SS metrics of class II in terms of single variable functions, and we obtain the constraints that the local thermal equilibrium condition imposes on these functions for the regular models.

Section \ref{sec-regular-0} is devoted to analyzing generic geometric, hydrodynamic and thermodynamic properties of the regular models. We obtain the metric line element, the expression for the {\em energy density} $\rho$ and the {\em pressure} $p$, as well as the expansion $\theta$ of the fluid velocity $u$. We also obtain their associated thermodynamics, namely, we determine the {\em specific entropy} $s$, the {\em matter density} $n$ and the {\em temperature} $\Theta$. Finally, we obtain an implicit expression for the square of the speed of sound $c_s^2$ in terms of the hydrodynamic quantities $\rho$ and $p$, $c_s^2 = \chi(\rho, p)$. 

In section \ref{sec-chi-pi} we obtain the regular models which are compatible with the equation of state of a generic ideal gas, $p=\tilde{k} n \Theta$, that is, with a speed of sound in the form $\chi = \chi(\pi)$, $\pi = p/\rho$ \cite{CFS-LTE}. We partially integrate the equations and show that two subfamilies must be distinguished.

In sections \ref{sec-ideal-parabolic} and \ref{sec-ideal-non} we consider these two subfamilies, the parabolic and the non-parabolic ideal regular models. In both cases we study the necessary conditions for physical reality by analyzing the energy condition and the relativistic compressibility conditions. Their associated thermodynamics are also outlined, and two distinguished thermodynamic schemes are considered in more detail: a generic ideal gas scheme and a thermodynamic scheme with the same (homogeneous) temperature as the $\gamma$-law models of the FLRW limit.

Finally, section \ref{sec-conclusions} is devoted to the summary and discussion of the results and to comment on further work currently underway.


\section{Towards a physical interpretation of the perfect fluid solutions}
\label{sec-physical}

Perfect fluid solutions of the Einstein equations have usually been obtained by imposing symmetries or other geometric properties on the gravitational field, or by requiring kinematic constraints \cite{Kramer}. A well-known problem in general relativity is the study of the possible physical interpretation of the known families of solutions. In this section we analyze this problem and present a method to answer it.

\subsection{Macroscopic necessary conditions for physical reality}
\label{subsec-necessary-conditions}

The evolution of a relativistic perfect fluid is described by an 
energy tensor in the form $T = (\rho+ p) u \otimes u + p \, g$, and submitted to the conservative condition: 
\be \label{ceq}
\hspace{-5mm} {\rm C} :  \qquad \qquad  \nabla \cdot [(\rho+ p) u \otimes u + p \, g] = 0 \, .
\ee
This constraint consists of a differential system of four equations on five {\em hydrodynamic quantities}  ({\em unit velocity} $u$, {\em energy density} $\rho$, and {\em
pressure} $p$). It is necessary to impose complementary physical requirements that (i) offer a causal closure\footnote{That is to say, the set of complementary equations to be added to the system in order that it admits unicity of the Cauchy problem.}  of this system, and (ii) guarantee that $T$ satisfies the general necessary constraints that a physical continuous medium is expected to meet.

 Pleba\'nski \cite{Plebanski} {\em energy conditions} are necessary algebraic conditions for physical reality and, in the perfect fluid case, they state: 
\begin{equation} \label{e-c}
\hspace{-5mm} {\rm E} :  \qquad \qquad  -\rho < p \leq \rho  \, .
\end{equation}
The determination of the spacetime regions where these constraints hold is a basic query in analyzing a given perfect fluid solution.

Furthermore, if we want to describe the evolution of a thermodynamic perfect fluid in {\em local thermal equilibrium}, the usual causal closure that we must add to the hydrodynamic quantities $\{u, \rho, p\}$ is a set $\{n, \epsilon, s, \Theta\}$ of {\em thermodynamic quantities} ({\em matter density} $n$, {\em specific internal energy} $\epsilon$, {\em temperature} $\Theta$, and {\em specific entropy} $s$) constrained by the common thermodynamic laws \cite{Eckart}. 
Namely, the conservation of matter (a dot denotes directional derivative with respect to $u$):
\begin{equation}  
\nabla \cdot (nu) = \dot{n} + n \theta = 0 \, ,  \label{c-masa}
\end{equation}
the {\em local thermal equilibrium relation} that can be written as:
\begin{equation}
\Theta \dif s = \dif h - \frac{1}{n} \dif p \, ,  \qquad h \equiv \frac{\rho+p}{n} \, , \label{re-termo}
\end{equation}
where $h$ is the {\em relativistic specific enthalpy}, and the decomposition defining the specific internal energy:
\begin{equation}
\rho= n(1+\epsilon) \, .  \label{masa-energia} 
\end{equation}

In \cite{Coll-Ferrando-termo}  (see also \cite{CFS-LTE}) we have shown that the causal closure \{(\ref{c-masa}),(\ref{re-termo}),(\ref{masa-energia})\} of the local thermal equilibrium admits a purely hydrodynamic formulation: {\em a non isoenergetic ($\dot{\rho} \not= 0$) perfect energy tensor $T$ evolves in local thermal equilibrium if, and only if, the hydrodynamic quantities $\{u, \rho, p\}$ fulfill the hydrodynamic sonic condition}: 
\begin{equation} \label{lte-chi}
\hspace{-5mm} {\rm S} :  \qquad \qquad     \dif \chi \wedge \dif p \wedge \dif \rho = 0 \, , \qquad \chi \equiv \frac{\dot{p}}{\dot{\rho}}   \, .
\end{equation}
Then, the {\em indicatrix of the local thermal equilibrium} $\chi$ is a function of state, $\chi = \chi(\rho,p)$, which physically represents the square of the {\em speed of sound} in the fluid, $\chi (\rho ,p) \equiv  c^2_{s}$.

When the conservation equation C and the hydrodynamic sonic condition S hold we say that $T\equiv \{u, \rho, p\}$ defines the {\em hydrodynamic flow} of a thermodynamic perfect fluid in local thermal equilibrium. Then, {\em thermodynamic schemes} $\{n, \epsilon, s, \Theta\}$ exist such that  $\{u, \rho, p, n, \epsilon, s, \Theta\}$ is a solution of the {\em fundamental system of the perfect fluid hydrodynamics} \{(\ref{ceq}),(\ref{c-masa}),(\ref{re-termo}),(\ref{masa-energia})\}.

Each thermodynamic scheme $\{n, \epsilon, s, \Theta\}$ associated with a hydrodynamic flow $T\equiv \{u, \rho, p\}$ is determined \cite{CFS-LTE} by a particular solution $s = s(\rho, p)$ to the equation $\dot{s}=0$ and a particular solution $n= n(\rho,p)$ to the equation (\ref{c-masa}). 

A basic physical requirement imposed on the thermodynamic schemes is the positivity of the matter density, of the temperature and of the specific internal energy,
\begin{equation} \label{P}
\hspace{-5mm} {\rm P} :  \qquad \qquad     \Theta > 0 \, , \qquad  \qquad   \rho > n > 0   \, .
\end{equation}

On the other hand, in order to obtain a coherent theory of shock waves for the fundamental system of perfect fluid hydrodynamics one must impose the relativistic compressibility conditions  \cite{Israel, Lichnero-1, Anile, Lichnero-2}:
\begin{equation}
\hspace{-5mm} {\rm H}_1 : \qquad \qquad    (\tau'_p)_s < 0 \, , \qquad \qquad (\tau''_p)_s > 0 \, , 
 \label{cc-1}
 \end{equation}
 \begin{equation}
\hspace{-5mm} {\rm H}_2 : \qquad \qquad   (\tau'_s)_p > 0 \, , 
\label{cc-2}
\end{equation}
where the function of state $\tau = \tau(p, s)$ is the {\em dynamic volume}, $\tau = \hat{h}/n$, $\hat{h} = h/c^2$ being the dimensionless enthalpy index.

In \cite{CFS-CC} we have shown that the compressibility conditions H$_1$ only restrict the hydrodynamic quantities, and that they can be stated in terms of the function of state $c_s^2 = \chi(\rho,p)$: 
\begin{equation}
\hspace{-5mm} {\rm H}_1 : \qquad \qquad 0 < \chi < 1 \, , \qquad    (\rho+p)(\chi \chi_{p}' + \chi_{\rho}') + 2 \chi(1-\chi) > 0   \, .       \label{cc-1-chi}
\end{equation}
However, compressibility condition H$_2$ imposes constraints on the thermodynamic scheme and it can be stated as \cite{CFS-CC}:
\be \label{H2-Theta}
\hspace{-5mm} {\rm H}_2 : \qquad \qquad  2 n \Theta > \frac{1}{s_{\rho}'}    \, .
\ee

The role of the general necessary macroscopic conditions C, E, S, P, H$_1$ and H$_2$ specified backward will be explained below.


\subsection{Procedure to determine physically admissible perfect fluid solutions }
\label{subsec-physical-solutions}

Einstein equations impose the conservation equation C on the space-time energy tensor $T$. In looking for perfect fluid solutions that satisfy the other physical requirements exposed in the subsection above, note that according to its nature we must distinguish two types of conditions: 
\begin{itemize}
\item[a)] 
{\em Hydrodynamic constraints}: the conservation equation C, the energy conditions E, the hydrodynamic sonic condition S, and the compressibility conditions H$_1$ exclusively involve the hydrodynamic variables $\{u, \rho, p\}$. They fully determine the hydrodynamic flow of the thermodynamic fluid in local thermal equilibrium and, consequently, restrict the admissible gravitational field as a consequence of the Einstein equations.
\item[b)]
{\em Thermodynamic constraints}: the positivity conditions P and the compressibility condition H$_2$ restrict the thermodynamic schemes $\{n, \epsilon, s, \Theta\}$ associated with a hy\-dro\-dynamic flow $\{u, \rho, p\}$. Consequently, they do not restrict the gravitational field and the admissible thermodynamics offer different physical interpretations for a given hydrodynamic perfect fluid flow. 
\end{itemize}

According to these considerations, we propose a method to analyze the possible physical reality of known relevant families of perfect fluid solutions.  The procedure follows five steps:
\begin{description}
\item[{\bf Step 1}]
Determine the subfamily of the thermodynamic solutions by imposing the hydrodynamic sonic condition S on the solutions to the conservative equations C. 
\item[{\bf Step 2}]
Obtain, for this subfamily, the coordinate dependence of the hydrodynamic quantities $u$, $\rho$, $p$, and  the indicatrix function $c_s^2 = \chi(\rho,p)$.
\item[{\bf Step 3}]
Analyze, for these thermodynamic solutions, the hydrodynamic constraints for physical reality, namely, the energy conditions E and the compressibility conditions H$_1$.
\item[{\bf Step 4}]
Obtain the thermodynamic schemes $\{n, \epsilon, s, \Theta\}$ associated with these solutions. The specific entropy $s$ and the matter density $n$ are of the form $s= s(\bar{s})$ and $n= \bar{n}R(\bar{s})$, where $s(\bar{s})$ and $R(\bar{s})$ are arbitrary real functions of a particular solution $\bar{s}=\bar{s}(\rho, p)$ to the equation $\dot{s}=0$, and $\bar{n}=\bar{n}(\rho,p)$ is a particular solutions to the equation (\ref{c-masa}) \cite{CFS-LTE}. Then, $\Theta$ and $\epsilon$ are given, respectively, by (\ref{re-termo}) and (\ref{masa-energia}).
\item[{\bf Step 5}]
Analyze, for the thermodynamic schemes $\{n, \epsilon, s, \Theta\}$ already obtained, the general thermodynamic constraints for physical reality, namely, the positivity conditions P and the compressibility condition H$_2$.
\end{description}
This is basically the same approach we have used elsewhere in studying the ideal gas Stephani universes \cite{CFS-CC, CF-Stephani}, the classical ideal gas solutions \cite{CFS-CIG} and the singular models of the thermodynamic class II Szekeres-Szafron solutions \cite{PSS}. In this paper we will apply it to study the regular models.


\subsection{Perfect fluid solutions modeling non-perfect fluids}
\label{subsec-physical-non-perfect}

The relativistic thermodynamic theory of irreversible processes was developed by Eckart \cite{Eckart} and is currently known as the {\em standard irreversible thermodynamics}. Israel and Stewart \cite{Israel-b, Israel-St} proposed an {\em extended irreversible thermodynamics} that solves the causal shortcomings of Eckart's theory. For a broad manual on extended thermodynamics see \cite{Jou-Casas}, and a comprehensive summary of both the standard and the extended approaches can be found in \cite{ReZa}.

According to the theory of thermodynamics of irreversible processes the transport coefficients of thermal conductivity, of shear-viscosity, and of bulk-viscosity play an important role. They appear in the constitutive equations linking dissipative fluxes (anisotropic pressures, bulk viscous pressure and energy flux) with the kinematic coefficients of fluid flow (shear, expansion and acceleration).  

The perfect fluid approximation can be considered when the transport coefficients of a fluid vanish (or are negligible). In this case, the energetic evolution of this fluid does not differ from that of a perfect fluid. Moreover, Eckart's thermodynamic theory reduces down to the fundamental system of the perfect fluid hydrodynamics exposed in subsection \ref{subsec-necessary-conditions} above.

A non-perfect fluid is a fluid with at least a non-zero transport coefficient. For this fluid, the energetic evolution is, generically, described by an energy tensor with energy flux and anisotropic pressures. However, when a non-perfect fluid admits particular evolutions in which the dissipative fluxes vanish, these evolutions are well described by a perfect energy tensor, and are usually called {\em equilibrium states} \cite{ReZa}. Moreover, all the thermodynamic relations of the perfect fluid hydrodynamics remain valid. Furthermore, the shear, the expansion and the acceleration of the fluid undergo strong restrictions as a consequence of the constitutive equations. For such equilibrium states \cite{ReZa}:
\begin{itemize}  
\item[-]
If the shear viscosity coefficient does not vanish, then the fluid shear vanishes.
\item[-]
If the bulk viscosity coefficient does not vanish, then the fluid expansion vanishes.
\item[-]
If the thermal conductivity coefficient does not vanish, then the fluid acceleration is constrained by the relation:
\be \label{Fourier}
a = - \perp \dif \ln \Theta \, , 
\ee
where $\perp$ denotes the orthogonal projection to the fluid velocity.
\end{itemize}

After these considerations it seems reasonable to look for perfect fluid solutions of the Einstein equations that describe both (i) a thermodynamic perfect fluid in local thermal equilibrium, and (ii)  a non-perfect fluid in equilibrium. In this paper we present some Szekeres-Szafron solutions that can describe an inviscid (with negligible shear and bulk viscosity coefficients) non-perfect fluid (with non-vanishing thermal conductivity).


\section{Class II Szekeres-Szafron metrics in local thermal equilibrium}
\label{sec-classII-termo}

The metric line element of the class II Szekeres-Szafron solutions takes the expression:
\begin{equation} \label{SS-canonica}
d s^2 = - dt^2 + \phi^2[(B + P)^2 d z^2 + C^2 (d x^2 + d y^2)] \, ,
\end{equation}
where 
\begin{eqnarray} \label{SS-metricfunctions}
\phi = \phi(t)  \,  , \qquad B=B(t,z) \,  , \qquad P= S \,C \, , \\[1mm]
C = C(x,y) \equiv [1 + \frac{k}{4} (x^2 + y^2)]^{-1}, \qquad k \equiv 0, 1, -1  , \label{II-C} \\[1mm] 
S = S(z,x,y) \equiv  \frac12 U(z) (x^2 + y^2) + V_1 (z) x +  V_2 (z) y + 2\,W(z) \,   .  \label{II-S}  
\end{eqnarray}
These metrics are perfect fluid solutions when the above metric functions fulfill:
\begin{equation} 
\ddot{B} + \frac{3 \, \dot{\phi}}{\phi} \dot{B} -  \frac{k}{\phi^2} B = \frac{1}{\phi^2}(U + kW) \, . \label{eq-SS-II}
\end{equation}
Moreover, the pressure $p$ and the energy density $\rho$ are given by: 
\begin{eqnarray} \label{pressure-II}
p = -\left[\frac{2\, \ddot{\phi}}{\phi} + \frac{\dot{\phi}^2}{\phi^2} + \frac{k}{\phi^2}\right]   \,  ,  \\[1mm]
\rho = \frac{3 \, \dot{\phi^2}}{ \phi^2} + \frac{3k}{\phi^2} + \frac{2 \, [\phi \, \dot{\phi}\, \dot{B} - k(B + W) -U]}{\phi^2 (B + P)}   \,  ,
\label{density-II}
\end{eqnarray}
and the unit velocity is $u= \partial_t$, it is geodesic and its expansion is:
\begin{equation} \label{expansion-II}
\theta = \frac{3 \, \dot{\phi}}{ \phi} + \frac{\dot{B}}{B + P}  \,  .
\end{equation}
Note that the dot denotes both partial derivative with respect $t$ and directional derivative with respect to $u$. 

Under the hypothesis $\rho + p \not=0$, the FLRW limit follows if, and only if, $\dot{B}=0$ (and then $k(B+W)+U=0$). We name \cite{FS-SS} {\em strict Szekeres-Szafron metrics} those of the form (\ref{SS-canonica}) with $\dot{B}\not=0$. 
A strict SS metric (\ref{SS-canonica}) admits a $G_3$ on $S_2$ if, and only if, $P= P(z)$ ($V_1 = V_2 =0$ and $U = k W$) \cite{FS-SS}.  
Finally, the non-conformally flat barotropic limit follows when, in addition to having a $G_3$ on $S_2$, the function $B+P$ factorizes and we obtain the canonical form of the Kantowski-Sachs metrics  and of their parabolic and hyperbolic counterparts \cite{Krasinski, K-S}.

The study of the differential equation (\ref{eq-SS-II}) leads to the following result:
\begin{lemma} \label{lemma-B}
The class II Szekeres-Szafron metrics take the expression (\ref{SS-canonica}), where 
\begin{equation} \label{B-ab}
\hspace{-1.5cm} B = \alpha a + \beta c +b \, , \qquad a=a(z) \, , \quad b=b(z) \, , \quad c = c(z) \equiv k(b+W) + U  ,
\ee
$\alpha = \alpha(t)$ and $\beta=\beta(t)$ being particular solutions, $\alpha \not= \alpha_0 (k \beta +1)$, $\alpha_0= constant$, of the linear differential equations
\be \label{eq-alpha-beta}
\ddot{\alpha} + \frac{3 \, \dot{\phi}}{\phi} \dot{\alpha} - \frac{k}{\phi^2} \alpha =0 \, ,    \qquad 
\ddot{\beta} + \frac{3 \, \dot{\phi}}{\phi} \dot{\beta} - \frac{1}{\phi^2} (k \beta +1) =0 \, . 
\ee
\end{lemma}
This lemma was presented without argument in \cite{cfERE}. A sketch of its proof is the following. If $\beta(t)$ is a solution of the second equation in (\ref{eq-alpha-beta}), it follows that $k \beta + 1$ is a solution of the first equation in (\ref{eq-alpha-beta}), and $B_0 = \beta(t) (U + kW)$ is a particular solution of the linear differential equation (\ref{eq-SS-II}). Thus, if $\alpha(t)$ is an independent particular solution of the first equation in (\ref{eq-alpha-beta}), $\alpha \not= \alpha_0 (k \beta +1)$, it follows that the general solution of the homogeneous equation associated with (\ref{eq-SS-II}) is $a(z) \alpha(t) + b(z)( k \beta + 1)$. Thus, if we add to this expression the particular solution $B_0$, we obtain the general solution to equation (\ref{eq-SS-II}).


\subsection{Thermodynamic constraints for Szekeres-Szafron solutions of class II}

Elsewhere \cite{FS-SS} we have proved that, for the {\em generic class II Szekeres-Szafron metrics} (non barotropic and expanding strict SS solutions), the hydrodynamic sonic condition (\ref{lte-chi}) is equivalent to $\dif \theta \wedge \dif \rho \wedge \dif t = 0$. From this constraint and the expressions (\ref{density-II}) of the energy density and (\ref{expansion-II}) of the expansion of a class II SS metric, we obtain (see proposition 1 in \cite{PSS}):
\begin{proposition} \label{lemma-termo}
A generic class II Szekeres-Szafron perfect fluid solution evolves in local thermal equilibrium if, and only if, it satisfies at least one of the following three conditions:
\begin{itemize}
\item[(i)] 
It admits a $G_3$ on $S_2$, that is, $P=P(z)$.
\item[(ii)] 
It is a singular model, that is, $k = U =0$.

\item[(iii)] 
It is a regular model, that is, $k^2 + U^2 \not =0$ and the metric functions fulfill 
\begin{equation} \label{lte-regular}
\partial_z \left[\frac{\dot{B}}{k(B+W)+U}\right] = 0 \,  .
\end{equation}
\end{itemize}
\end{proposition}
Note that the three three cases in proposition  \ref{lemma-termo} define three non-disjoint families of solutions in local thermal equilibrium. Anyway, the entire study of the thermodynamic SS solutions of class II involves analyzing these three cases. In \cite{PSS} we have studied in depth the singular models by obtaining their associated thermodynamic schemes and by outlining some relevant physical models. In this paper we tackle a similar study for the regular models.

First we analyze equation (\ref{lte-regular}) for $B$ of the form (\ref{B-ab}). Note that $E \equiv k(B+W)+U = k \alpha a + (k \beta +1) c$. Then, a straightforward calculation shows that, under the constraint $\alpha \not= \alpha_0 (k \beta +1)$, condition (\ref{lte-regular}) holds if, and only if, $a =\tilde{c} \, c$, $\tilde{c} = constant$. Then, $B= (\beta + \tilde{c} \, \alpha) c + b$, with $\beta + \tilde{c} \, \alpha$ a solution of the second equation in (\ref{eq-alpha-beta}). Thus, we obtain the following result that was presented without proof in \cite{cfERE}:
\begin{proposition}  \label{propo-a=0}
The metric of a regular class II Szekeres-Szafron model takes the expression (\ref{SS-canonica}), where 
\begin{equation}
B =  \beta c +b \, , \qquad  b=b(z) \, , \quad c = c(z) \equiv k(b+W) + U  ,
\ee
$\beta=\beta(t)$ being a particular solution of the linear differential equation:
\be \label{eq-beta}
\ddot{\beta} + \frac{3 \, \dot{\phi}}{\phi} \dot{\beta} - \frac{1}{\phi^2} (k \beta +1) =0 \, . 
\ee
\end{proposition}
%


\section{Regular models}
\label{sec-regular-0}


\subsection{Metric and hydrodynamic variables: energy density and pressure} 
\label{subsec-metric-regular}

We can simplify the expression of the metric of a regular  class II SS solution by writing $c(z) = \varepsilon f(z)$, $\varepsilon = 0, 1$, and $Q = (b + P)/f$, and by changing the coordinate $z$ as $d \tilde{z} = f(z) d z$. Then, reusing the notation $z$ for the new $\tilde{z}$ and redefining new functions $U$, $W$, $V_1$ and $V_2$ from the old ones as $(U+ k b/2)/f$, $(W+b/2)/f$,  $V_1/f$ and $V_2/f$, we obtain the following {\em canonical form of the regular models}:
\begin{equation} \label{regular-SS-canonica}
d s^2 = - dt^2 + \phi^2[(\varepsilon \beta + Q)^2 d z^2 + C^2(d x^2 + d y^2)] \, ,
\end{equation}
where $\phi = \phi(t)$ and $\beta=\beta(t)$ are constrained by equation (\ref{eq-beta}), and $Q = SC$, where $C$ and $S$ take the form (\ref{II-C}) and (\ref{II-S}), with $U + k W= \varepsilon$.\\
The {\em pressure} $p$ and the {\em energy density} $\rho$ are then given by: 
\begin{eqnarray} \label{pressure-regular}
p = -\left[\frac{2\, \ddot{\phi}}{\phi} + \frac{\dot{\phi}^2}{\phi^2} + \frac{k}{\phi^2}\right]   \,  ,  \\[1mm]
\rho = \frac{3 \, \dot{\phi^2}}{ \phi^2} + \frac{3k}{\phi^2} + \frac{2 \, \varepsilon [\phi \, \dot{\phi}\, \dot{\beta} - (k \beta + 1)]}{\phi^2 (\varepsilon \beta + Q)}   \,  ,
\label{density-regular}
\end{eqnarray}
And the {\em expansion} of the fluid is:
\begin{equation} \label{expansion-parabolic}
\theta = \frac{3 \, \dot{\phi}}{ \phi} + \frac{\varepsilon \dot{\beta}}{\varepsilon \beta + Q} = \partial_t [\ln\{\phi^3( \varepsilon \beta + Q)\}] \,  .
\end{equation}
The regular models (\ref{regular-SS-canonica}) depend on an arbitrary function of time ($\phi(t)$ and $\beta(t)$ are submitted to the equation (\ref{eq-beta})) and three arbitrary functions, $V_1 (z)$, $V_2 (z)$ and $W(z)$, of the coordinate $z$. Now, we recover the FLRW limit by making $\varepsilon=0$. Note that the metric canonical form (\ref{regular-SS-canonica}) prevents metrics admitting a $G_3$ on flat two-dimensional orbits ($k=0$) because $V_1 = V_2 =0$ and $U=kW$ implies $\varepsilon = U + k W = 0$.
%


\subsection{Thermodynamic scheme: entropy, matter density and temperature} 
\label{subsec-scheme-regular}

We know that the regular metrics (\ref{regular-SS-canonica}) define perfect fluid solutions in l.t.e., and the hydrodynamic variables pressure and energy density are given in (\ref{pressure-regular}) and (\ref{density-regular}), respectively. Now we shall solve the inverse problem \cite{CFS-LTE} for the perfect fluid solutions (\ref{regular-SS-canonica}). Namely, we shall obtain the full set of thermodynamic quantities: specific entropy $s$, matter density $n$ and temperature $\Theta$. 

For the sake of clarity, from now on we compress the expression of the hydrodynamic variables by defining two functions $\tau$ and $\xi$ that depend on time: 
\begin{eqnarray} \label{pressure-regular-tau}
p = - 2 \frac{\ddot{\phi}}{\phi} - \tau  \,   , \qquad  \qquad \quad  \tau =\tau(t) \equiv  \frac{\dot{\phi}^2}{\phi^2} + \frac{k}{\phi^2} \, , \\[1mm]
\rho = 3 \tau  -  \frac{2 \, \varepsilon \, \dot{\xi}}{\phi^2 (\varepsilon \beta + Q)}\,  ,  \qquad \xi = \xi(t) \equiv \phi^2 \dot{\beta} \,    .
\label{density-regular-tau}
\end{eqnarray}
The last expression follows by taking into account that, with the above definition of $\xi(t)$, equation (\ref{eq-beta}) is equivalent to:
\be \label{eq-beta-gamma}
\dot{\xi}  \equiv (\phi^2 \dot{\beta})^{\cdot}  = -\phi \dot{\phi} \dot{\beta} + k\beta+1    \, . 
\ee

For the strict SS metrics ($\varepsilon= 1$), we can isolate the function $Q$ from (\ref{density-regular-tau}), and we obtain:
\be \label{Q-rho-p}
Q =  \left[\frac{2 \dot{\xi}}{\phi^2 (3 \tau - \rho)} - \beta \right] \equiv Q(\rho,p) \, .
\ee
The expressions (\ref{pressure-regular-tau}), (\ref{density-regular-tau}) and (\ref{Q-rho-p}) are formally identical to those obtained in \cite{PSS} for the singular models if we take $\xi=- 1/\phi$ and $k=0$. Thus, with a similar reasoning to that presented in \cite{PSS} we can obtain the general expression of the thermodynamic quantities $n$, $s$ and $\Theta$, which brings us to the following:
\begin{proposition} \label{prop-s-n-regular}
The thermodynamic schemes associated with the regular models (\ref{regular-SS-canonica}) are determined by two arbitrary functions $s(Q)$, $s'(Q) \not=0$, and $r(Q)\not=0$. The specific entropy $s$, the matter density $n$, and the temperature $\Theta$ are given by:
\begin{eqnarray}  \label{s-n-regular}
s = s(Q) \equiv s(\rho, p) \, ; \qquad  \ n = \frac{1}{\phi^3 (\beta + Q)r(Q)} \equiv n(\rho,p) \, ; \\[1mm]
\label{T-regular}
\Theta = \ell(Q) \lambda(t) + m(Q) \mu(t) \equiv \Theta(\rho,p)  \, , \\[1mm]
\label{T-regular-b}
 \ell(Q) \equiv \frac{r'}{s'} \, , \qquad  m(Q) \equiv \frac{1}{s'}[Q r' + r]    \, , \\[1mm]
\label{lambda-mu}
\lambda(t) \equiv - 2 \phi \dot{\xi} + \beta  \phi^3(p+ 3\tau)  \, , \quad \quad  \mu(t) \equiv \phi^3(p+ 3\tau)  \, .
\end{eqnarray}
\end{proposition}
Note that we also use the prime for the derivative of any function depending on a sole variable.


\subsection{The indicatrix function: speed of sound}
\label{subsec-chi-regular}

The square of the speed of sound in terms of the hydrodynamic variables, $c_s^2 = \chi(\rho,p)$, is given by the indicatrix function (\ref{lte-chi}). From (\ref{density-regular-tau}) we can compute $\dot{\rho}$ by taking into account (\ref{pressure-regular-tau}), (\ref{eq-beta-gamma}) and (\ref{Q-rho-p}), and we obtain:
\begin{eqnarray} \label{rho-punt}
\dot{\rho} = \tilde{A} \rho^2 + \tilde{B}  \rho + \tilde{C}  \, , \\ 
\tilde{A}  \equiv \frac{\xi}{2\dot{\xi}}  \, , \quad  \tilde{B}  \equiv  - \frac{3\dot{\phi}}{\phi}+ (p-3 \tau)\tilde{A}  \, , \quad  \tilde{C}  \equiv  - \frac{3\dot{\phi}}{\phi}p -3 \tau p \tilde{A}  \, . \label{rho-punt-b}
\end{eqnarray}
Consequently, we have the following result:
\begin{proposition} \label{prop-chi-regular}
For the regular models (\ref{regular-SS-canonica}), the square of the speed of sound takes the expression:
\be  \label{chi-regular}
c_s^2  = \chi(\rho,p)  \equiv \frac{1}{{\cal A}(p) \rho^2 + {\cal B}(p) \rho + {\cal C}(p)} \, ,
\ee
where ${\cal A}$, ${\cal B}$ and ${\cal C}$ are the functions of $t$ (and then of $p$) given by:
\begin{equation} 
\hspace{-12mm} {\cal A}(p) \equiv  \frac{\xi}{2\dot{\xi} \dot{p}} \, , \qquad {\cal B}(p) \equiv  - \frac{3\dot{\phi}}{\phi\dot{p}}+ (p-3 \tau){\cal A} \, ,  \qquad {\cal C}(p) \equiv   - \frac{3\dot{\phi}}{\phi \dot{p}}p -3 \tau p {\cal A}  \label{ABCcal}  \, .   
\end{equation}
\end{proposition}
Note that (\ref{chi-regular}) provides an expression of the indicatrix function which is implicit in the variable $p$. For a specific solution $(\phi(t), \beta(t))$ of the equation (\ref{eq-beta}) we can obtain $p(t)$ from (\ref{pressure-regular-tau}), and thus we get $t(p)$. Then, the explicit form of $\chi(\rho,p)$ can be obtained (see forthcoming sections).

In this section we have achieved steps 1, 2 and 4 of the procedure proposed in subsection \ref{subsec-physical-solutions}. We could formally impose the restrictions required in step 3 (energy and compressibility conditions H$_1$), and we would get inequalities involving third order derivatives of the generic metric functions. Nevertheless, it may be more practical to make this study for a subclass of solutions that fulfills complementary physical requirements, and once we have obtained the explicit form of $\chi(\rho,p)$. In the following sections we consider the solutions that have the hydrodynamic properties of a generic ideal gas.


\section{Models with the hydrodynamic behavior of a generic ideal gas}
\label{sec-chi-pi}


Now we will analyze when the regular models considered above are compatible with the equation of state of a {\em generic ideal gas}, namely:
\begin{equation}
p = \tilde{k} n \Theta  \, , \qquad \quad    \tilde{k} \equiv {k_B \over m} \,  .  \label{gas-ideal}
\end{equation}
In \cite{CFS-LTE} we have solved the direct problem for the generic ideal gases by studying the hydrodynamic constraints that equation (\ref{gas-ideal}) imposes, and in \cite{CFS-CC} we have analyzed the compressibility condition H$_1$ for this particular case. These results can be summarized in the following.

\begin{lemma}
A perfect energy tensor $T=\{u,\rho,p\}$ represents the evolution of a generic ideal gas in l.t.e. if, and only if, it fulfills the {\em ideal gas sonic condition}:
\be \label{chi-gas-ideal}
\hspace{-5mm} {\rm S^{\rm G}} :  \qquad \quad     \chi = \chi(\pi) \not= \pi \, , \qquad \chi = \frac{\dot{p}}{\dot{\rho}} \, , \qquad \pi = \frac{p}{\rho} \, .
\ee
For a fluid with $\chi=\chi(\pi)$, compressibility conditions ${\rm H}_1$ given in (\ref{cc-1-chi}) are equivalent to:
\begin{equation}
\hspace{-5mm} {\rm H}_1^{\rm G} :  \qquad \quad       0 < \chi < 1 \, , \qquad   \zeta \equiv (1+\pi)(\chi-\pi) \chi'  + 2 \chi(1-\chi) > 0   \, .       \label{cc-ideal}
\end{equation}
\end{lemma}

On the other hand, although there are continuous media with negative pressure, the equation of sate (\ref{gas-ideal}) and the positivity conditions P given in (\ref{P}) imply a non-negative thermodynamic pressure, $p \geq 0$. Moreover here we shall consider non-shift perfect fluids ($\rho \not=p$). Consequently, the energy conditions E given in (\ref{e-c}) become: 
\begin{equation} \label{e-c-G}
\hspace{-5mm} {\rm E}^{\rm G} : \qquad \qquad    \rho > 0 \, , \qquad  0 \leq \pi < 1 \, , \quad  \pi = \frac{p}{\rho} \, .
\end{equation}

Then, in order to study the solutions with the hydrodynamic behavior of a generic ideal gas, we can 
fairly modify the procedure exposed in subsection \ref{subsec-physical-solutions}:
\begin{description}
\item[{\bf Step 1}']
Determine the subfamily of the ideal gas solutions by imposing the ideal gas sonic condition S$^{\rm G}$. 
\item[{\bf Step 2}']
Obtain, for this subfamily, the coordinate dependence of the hydrodynamic quantities $u$, $\rho$, $p$, and  the indicatrix function $c_s^2 = \chi(\pi)$.
\item[{\bf Step 3}']
Analyze, for the ideal gas solutions, the hydrodynamic general constraints for physical reality, namely, the energy conditions E$^{\rm G}$ and the compressibility conditions H$_1^{\rm G}$.
\item[{\bf Step 4}']
Obtain the thermodynamic schemes $\{n, \epsilon, s, \Theta\}$ associated with the ideal gas solutions.
\item[{\bf Step 5}']
Analyze, for some physically relevant thermodynamic schemes $\{n, \epsilon, s, \Theta\}$, the thermodynamic general constraints for physical reality, namely, the positivity conditions P and the compressibility condition H$_2$.
\end{description}

\subsection{Study of the ideal sonic condition {\rm S}$^{\rm G}$}
\label{subsec-ideal-equations}

From the expression of the indicatrix function (\ref{chi-regular}) we obtain that, for the regular models, the ideal gas sonic condition S$^{\rm G}$ given in (\ref{chi-gas-ideal}) is equivalent to:
\be
{\cal A} p^2 = c_1 \, , \qquad {\cal B} p = c_2 \, , \qquad {\cal C} = c_3 \, , \qquad c_i = constant \, .
\ee
Then, if we use the expressions (\ref{ABCcal}) and we take into account that $\dot{\beta}\not=0$, we obtain:
\begin{lemma}
The regular models with an indicatrix function of the form (\ref{chi-gas-ideal}) fulfill the equations: 
\begin{equation} 
\hspace{-17mm} c_1 \equiv   \frac{\xi p^2}{2\dot{\xi} \dot{p}} \not=0  , \qquad c_2 \equiv  - 3\frac{\dot{\phi}}{\phi}\frac{p}{\dot{p}}+ c_1 \! \left(\!1-3 \frac{\tau}{p}\right)  ,  \qquad c_3 \equiv  - 3\frac{\dot{\phi}}{\phi}\frac{p}{\dot{p}}- 3 c_1 \frac{\tau}{p}.  \label{GI-equations}     
\end{equation}
where $c_i$ are constants.
\end{lemma}
From the expressions (\ref{pressure-regular-tau}) and (\ref{density-regular-tau}) of $p$, $\tau$ and $\xi$, it follows that equations (\ref{GI-equations}) and (\ref{eq-beta-gamma}) constitute a third-order differential system for the metric functions $\phi(t)$ and $\beta(t)$.

The second and the third equations in (\ref{GI-equations}) are compatible and equivalent when
\be
c_2 = c_1 + c_3 \, . \label{c_i-res}
\ee
Then, the third equation in (\ref{GI-equations}), which only involves  the metric function $\phi$, can be written as
\be \label{p-dot}
(c_3 \, p + 3 c_1 \tau) \, \dot{p} = - 3 \, \frac{\dot{\phi}}{\phi}\, p^2 \, ,
\ee
where $p$ and $\tau$ are given in (\ref{pressure-regular-tau}). The other two equations involve both $\phi$ and $\beta$ and they take the expression:
\begin{eqnarray}  \label{xi-punt-1}
\dot{\xi} = -\frac{\dot{\phi}}{\phi} \, \xi + k \beta + 1 \, , \qquad   \xi = \phi^2 \dot{\beta}  \, ,   \\
2 c_1\, \dot{\xi} =\left(\frac{p^2}{\dot{p}}\right) \xi \, .  \label{xi-punt-2}
\end{eqnarray}
A first accurate analysis of the equations leads to the following result, which will be proved in \ref{Ap-equations}:
\begin{lemma} \label{lemma-equations}
If the differential system \{(\ref{pressure-regular-tau}),(\ref{p-dot}),(\ref{xi-punt-1}),(\ref{xi-punt-2})\} admits a solution then at least one of the following conditions holds:
\begin{itemize}
\item[i)]
$\ddot{\phi} = 0 \, .$
\item[ii)]
$k=0$ and $\tau = c_0 \, p , \ c_0 = constant$.
\item[iii)]
$c_1= 2, \ c_3=6$ and $p=-2\tau$. 
\item[iv)]
$c_1= 1/8, \ c_3=3/8$ and $p= \tau$. 
\end{itemize}
%
\end{lemma}
In \ref{Ap-rarets} we show that cases (i) and (iii) are of limited physical interest: the first one has, necessarily, a negative pressure, and the second one leads to a negative energy density.  In following sections we focus on cases (ii) and (iv), and we study in detail the five steps required for analyzing the physical reality of the solutions. 


\section{Ideal parabolic regular models}
\label{sec-ideal-parabolic}

\subsection{Metric and hydrodynamic variables: energy density, pressure and speed of sound} 
\label{subsec-metric-parabolic-ideal}

In case (ii) we have a parabolic model ($k=0$) with $\tau = c_0 \, p$. Then, equations (\ref{pressure-regular-tau}) and (\ref{p-dot}) for the functions $\phi(t)$ and $p(t)$ are equivalent to:
\be  \label{phi-punt}
p = 3 \kappa^2 (\gamma-1)  \phi^{-3 \gamma}    \,   , \qquad \quad   \dot{\phi}  = \kappa \, \phi^{1-3 \gamma/2} \,  \, ,
\ee
where $\kappa$ is an arbitrary constant and $\gamma \equiv 1/(3c_0) +1$. Moreover, the constants $c_i$ are constrained by $c_3+3 c_1 c_0 = \gamma^{-1}$.

If we take into account (\ref{phi-punt}), equations (\ref{xi-punt-1}) and (\ref{xi-punt-2}) for the function $\beta(t)$ admit a solution if $3 \gamma^2 c_1 + \gamma -1 = 0$, and then:
\be \label{beta-0}
\beta = \beta(\phi) \equiv
\cases{
\beta_0 + \frac{2 \, }{\kappa^2 (9 \gamma^2 - 4)} \phi^{3 \gamma -2}  \, , \qquad {\rm if} \quad \gamma \not= 2/3  \cr
\beta_0 + \frac{1}{2 \kappa^2} \,  \ln \phi \, , \qquad \quad \qquad  \ \ \,   {\rm if} \quad  \gamma = 2/3
}
\ee
In addition, we obtain $\dot{\xi}= 3 \gamma/(2+3\gamma)$. Note that the function $\beta$ appears in the metric expression (\ref{regular-SS-canonica}) through $\epsilon \beta + Q$. Thus we can redefine $W(z)$ such that we can take $\beta_0 = 0$. On the other hand, for non-dust solutions with positive pressure, expression (\ref{phi-punt}) implies $\gamma > 1$. Moreover, the FLRW limit $\varepsilon=0$ leads to a barotropic evolution of the form $p=(\gamma -1) \rho$. These FLRW models fulfill the energy condition (\ref{e-c}) when $\gamma < 2$, and they are the so-called $\gamma$-law models \cite{Assad-Lima}. Hereinafter, the solutions with $\varepsilon = 1$ and $1 <\gamma <  2$ will be named {\em ideal parabolic regular models}. We can integrate the second equation in (\ref{phi-punt}), and considering (\ref{density-regular}) and (\ref{beta-0}), we obtain:
\begin{proposition} \label{prop-regular-parabolic-ideal}
The ideal parabolic regular models have a metric line element of the form 
\begin{equation} \label{regular-parabolic-SS-canonica}
\dif s^2 = - \dif t^2 + \phi^2[(\varepsilon \beta + Q)^2 \dif z^2 + \dif x^2 + \dif y^2] \, ,
\end{equation}
where $Q= Q(x,y,z)$, and $\phi(t)$ and $\beta(t)$ are given, respectively, by:
\begin{equation} \label{parabolic-SS-metricfunctions}
Q =  \frac{\varepsilon}{2}(x^2 + y^2) +V_1 (z) x +  V_2 (z) y + 2\,W(z) \,   ,   
\end{equation}
\begin{equation} \label{phi-t}
\phi(t) = \left[\frac32 \kappa \gamma \,(t-t_0) \right]^{\frac{2}{3 \gamma}}   ,  \qquad  \beta = \beta(\phi) \equiv \frac{2 \, }{\kappa^2 (9 \gamma^2 - 4)} \phi^{3 \gamma -2} .
\end{equation}
Moreover the pressure $p$ and the energy density $\rho$ are:
\begin{eqnarray} \label{pressure-parabolic-ideal}
p = \frac{3 \kappa^2 (\gamma-1) }{\phi^{3 \gamma}}    \,  ,  \\[1mm]
\rho = \frac{3 \kappa^2 }{\phi^{3 \gamma}} 
\left[1 - \frac{2 \varepsilon  \gamma (3 \gamma -2)}{ \kappa^2 (9 \gamma^2 - 4 ) \phi^{2- 3\gamma} \, Q + 2 \varepsilon}\right]  \, .
\label{density-parabolic-ideal}
\end{eqnarray}
And the speed of sound is given by: 
\be \label{chi-parabolic-ideal}
c_s^2  = \chi(\pi)  \equiv \frac{3 \, \gamma^2 \, \pi^2}{ (\pi + 1) [(3 \gamma + 1) \pi - (\gamma-1)]} \, , \qquad \pi \equiv \frac{p}{\rho} \, .
\ee
\end{proposition}
%


\subsection{Analysis of the solutions. Energy conditions} 
\label{subsec-energy-c}

It is worth remarking the following qualities of the solutions in proposition \ref{prop-regular-parabolic-ideal}:
\begin{itemize}
\item[(i)] 
The metric depends on three arbitrary functions of $z$, $V_1 (z)$, $V_2 (z)$ and $W(z)$, and two effective parameters, $\kappa$ and $\gamma$. The constant $t_0$ only determines an origin of time and it does not affect the metric.
\item[(ii)] 
Expression (\ref{phi-punt}) shows that the sign of the {\em amplitude parameter} $\kappa$ gives the sign of $\dot{\phi}$,  and thus it is positive in expanding models. Its square $\kappa^2$ determines the strength of the density $\rho$ and the pressure $p$. The {\em thermodynamic parameter} $\gamma$ is the only one that affects the equation of state (\ref{chi-parabolic-ideal}).
\item[(iii)] 
The only solutions admitting a $G_3$ on two-dimensional orbits are the FLRW models ($\varepsilon =0$).
\item[(iv)] 
In expanding models the solutions approach to the FLRW $\gamma$-law models for early times ($\phi \to 0$). In contracting models the solutions evolve to the FLRW $\gamma$-models.
\item[(v)] 
The solutions belong to the family of metrics considered by Szafron and Wainwright \cite{Szafron-Wain}, which was the first generalization with non-vanishing pressure of the Szekeres dust solutions.
\end{itemize}

Now we achieve the analysis of the energy conditions E$^{\rm G}$ by obtaining the spacetime domains where these conditions (\ref{e-c-G}) are fulfilled. From the expressions (\ref{pressure-parabolic-ideal}) and (\ref{density-parabolic-ideal}) for the pressure and density we obtain ($\varepsilon = 1$):
\be \label{ec-parabolic-ideal-0}
\pi^{-1}  = \frac{\rho}{p} = \frac{1}{\gamma-1}\left[1- \frac{2 \gamma (3 \gamma -2)}{Y+2}\right] \, , \qquad Y \equiv   \kappa^2 (9 \gamma^2 - 4) \phi^{2-3\gamma} Q \, .
\ee
Consequently, $\rho > p$ if either $Y < -2$ or $Y > \frac{(3 \gamma+2)(\gamma-1)}{2-\gamma}$, so that:
\begin{proposition} \label{prop-ec-parabolic-ideal}
The ideal parabolic regular models in proposition \ref{prop-regular-parabolic-ideal} fulfill the energy conditions {\rm E}$^{\rm G}$ given in (\ref{e-c-G}) in the spacetime domains where one of the following two conditions holds:
\be \label{ec-parabolic-ideal}
\kappa^2 \frac{(3 \gamma - 2)(2-\gamma)}{\gamma-1} \, Q > \phi^{3 \gamma-2} \, , \qquad \quad  \kappa^2 (9 \gamma^2 - 4) \, Q < -2 \phi^{3 \gamma-2} \, .
\ee
\end{proposition} 
Note that, for contracting models ($\kappa <0$), the spatial domain where conditions (\ref{ec-parabolic-ideal}) hold increases with time.

%


\subsection{Compressibility conditions} 
\label{subsec-compress-c-parabolic-ideal}

Now we will complete step 3' for the ideal parabolic regular models by analyzing the compressibility conditions H$_1^{\rm G}$.  From the expression (\ref{chi-parabolic-ideal}), a straightforward calculation leads to:
\be \label{chi-prima}
\hspace{-2cm} \chi' (\pi) = \frac{6 \, \gamma^2 \, \pi [(\gamma+1) \pi - (\gamma-1)]}{(\pi + 1)^2 [(3\gamma + 1) \pi - (\gamma-1)]^2}  \, ,  \quad \chi(0) =0 \, , \quad \chi(1) = \frac{3 \gamma^2}{4(\gamma + 1)} \, .
\ee
Thus, for any $\gamma \in ]1, 2[$, $\chi(\pi)$ has a vertical asymptote in $\pi_{\infty} = (\gamma-1)/(3 \gamma+1)$,
and it is negative and decreasing  in $[0 , \pi_{\infty}[$. Moreover, $\chi(\pi)$ reaches a relative minimum in $\hat{\pi}= (\gamma-1)/(\gamma+1)> \pi_{\infty}$. Then, it is easy to show that the first compressibility condition H$_1^{\rm G}$ in (\ref{cc-ideal}) holds in the interval, $\pi \in [\pi_m, 1[$, where 
\be \label{pi_m}
\pi_m = \frac{\gamma-1}{\gamma+1 + \sqrt{7 \gamma^2-3 \gamma^3}} \, , \qquad \pi_{\infty} < \pi_m < \hat{\pi} \, .
\ee

On the other hand, if we use (\ref{chi-parabolic-ideal}) and (\ref{chi-prima}) to replace $\chi$ and $\chi'$ in the second expression in (\ref{cc-ideal}) we obtain (for $\pi<1$):
\be
\zeta  = \frac{2 \, \gamma^2 \, \pi^3 (1- \pi)[(5 \gamma -2) \pi + (3 \gamma+2)(\gamma-1)]}{ (\pi + 1)^2(\pi+\gamma-1)^3} > 0 \, , 
\ee
Therefore, the second compressibility condition H$_1^{\rm G}$ holds provided that the first one does. Consequently, we have shown:
\begin{proposition} \label{prop-cc-parabolic-ideal}
The ideal parabolic  regular models in proposition \ref{prop-regular-parabolic-ideal} fulfill the compressibility conditions ${\rm H}_1^{\rm G}$  in the interval $]\pi_m , 1[$, where $\pi_m$ is given in (\ref{pi_m}).
\end{proposition} 
%


\subsection{Thermodynamic schemes: entropy, matter density and temperature} 
\label{subsec-scheme-parabolic-ideal}

Now we will perform step 4' by studying the full set of thermodynamics associated with the (strict, $\varepsilon =1$) ideal parabolic singular models. We must particularize the thermodynamic schemes presented in subsection \ref{subsec-scheme-regular} for the solutions in proposition \ref{prop-regular-parabolic-ideal}. Note that, from (\ref{phi-t}) and (\ref{pressure-parabolic-ideal}), we obtain
\be \label{phi-p}
\hspace{-20mm} \phi = \phi(p) \equiv  \left[\frac{3 \kappa^2(\gamma-1)}{p}\right]^{\frac{1}{3 \gamma}}, \qquad \beta = \beta(p) \equiv \frac{2}{\kappa^2 (9 \gamma^2-4)} \left[\frac{3 \kappa^2(\gamma-1)}{p}\right]^{1-\frac{2}{3 \gamma}}  . 
\ee
Then, taking into account (\ref{phi-p}) and (\ref{density-parabolic-ideal}) we can obtain :
\be \label{Q-alpha-rho-p-1}
 \phi^2 ( \beta + Q) =  - \frac{\tilde{K}}{\rho(\gamma-1) - p} \, , \qquad \tilde{K}  \equiv \frac{6 \gamma(\gamma-1)}{2+3\gamma}  > 0  \,   .
\ee
Consequently, we can determine an explicit expression for the function $Q(\rho,p)$:  
\be \label{Q-rho-p-1}
\hspace{-22mm} Q = Q(\rho,p) \equiv  - \hat{K}    \frac{\rho +(3 \gamma +1) p}{p^{1-\frac{2}{3 \gamma}}[\rho(\gamma-1) - p]}  , \qquad \hat{K} \equiv \frac{6 (\gamma-1)^2}{(9 \gamma^2-4)[3 \kappa^2(\gamma-1)]^{\frac{2}{3\gamma}}} > 0  .  
\ee

On the other hand, the functions $\lambda(t)$ and $\mu(t)$ defined in (\ref{lambda-mu}) can be computed in terms of $\phi$ and in terms of $p$ by using the expressions above:
\begin{eqnarray} \label{lambda-p}
\hspace{-1cm} \lambda =  -\frac{18 \gamma (\gamma-1)}{9 \gamma^2-4} \phi = c_{\lambda}  \, p^{-\frac{1}{3\gamma}} \,  ,  \qquad c_{\lambda} \equiv -\frac{18 \gamma (\gamma-1)}{9 \gamma^2-4}[3 \kappa^2(\gamma-1)]^{\frac{1}{3\gamma}}  < 0 , \\
\hspace{-1cm}  \mu =  3 \kappa^2 \gamma \phi^{-3 (\gamma-1)}= c_{\mu} \, p^{1- \frac{1}{\gamma}}  , \quad \qquad   c_{\mu} \equiv \frac{\gamma}{\gamma-1}[3 \kappa^2(\gamma-1)]^{\frac{1}{\gamma}}  > 0  \, . \label{mu-p}
\end{eqnarray}
Then, substituting all these formulas in expressions (\ref{s-n-regular}), (\ref{T-regular}) and (\ref{T-regular-b}) for $n$, $s$ and $\Theta$, we obtain:
\begin{proposition} \label{prop-scheme-parabolic-ideal}
The thermodynamics associated with the ideal parabolic regular models given in proposition \ref{prop-regular-parabolic-ideal} are determined by a specific entropy $s$ and a matter density $n$ of the form:
\be  \label{s-n-parabolic-ideal}
\hspace{-2cm} s = s(Q) \equiv s(\rho, p) \, , \quad  n =  \frac{[\rho(\gamma-1) - p] \, p^{\frac{1}{3 \gamma}} }{K \, r(Q) } \equiv n(\rho,p) , \quad    K \equiv \! \Big( \!\gamma - \frac23 \Big)  c_{\lambda}  < 0  \, ,  
\ee
where $s(Q)$, $s'(Q)\not=0$, and $r(Q)\not=0$ are arbitrary real functions of the function of state $Q=Q(\rho,p)$ given in (\ref{Q-rho-p-1}). Moreover the temperature is of the form (\ref{T-regular}), where $\lambda(p)$ and $ \mu(p)$ are given in  (\ref{lambda-p}) and (\ref{mu-p}), and  $\ell(Q)$ and $m(Q)$ are given in (\ref{T-regular-b}).
\end{proposition}


\subsection{Models with a generic ideal gas thermodynamic scheme} 
\label{subsec-scheme-parabolic-ideal-ideal}

When the hydrodynamic quantities $(u, \rho,p)$ fulfill the ideal gas sonic condition S$^{\rm G}$ given in  (\ref{chi-gas-ideal}), a thermodynamic scheme modeling a generic ideal gas in l.t.e. exists. Obtaining this scheme solves the {\em restricted inverse problem} for the indicatrix function $\chi = \chi(\pi)$, a problem that was previously studied (see lemma 4 in \cite{CFS-LTE}). In \cite{PSS} we have applied this study to the ideal singular models. A similar analysis for the ideal parabolic regular models leads to a {\em specific energy density}:
\be
e = \frac{\rho}{n} = e(\pi) \equiv e_0 \{[(3 \gamma +1) \pi + 1][(\gamma-1)-\pi]^{3(\gamma-1)}\}^{-\frac{1}{3\gamma-2}} ,   \label{e-pi}
\ee
and the other thermodynamic quantities are given by:
\begin{proposition}
The matter density $n$, the specific entropy $s$ and the temperature $\Theta$ of the generic ideal gas scheme associated with an ideal parabolic regular model take the expressions: 
\begin{eqnarray}
n(\rho,p) = \, \frac{1}{e_0} \{[(3 \gamma +1) p + \rho][(\gamma-1) \rho- p]^{3(\gamma-1)}\}^{\frac{1}{3\gamma-2}}  , \label{n-idealgas}  \\
s(\rho,p) = s_0 + \tilde{k} \ln \!  \left\{\! \frac{1}{p} \left[\frac{\rho+ (3 \gamma +1)p}{\rho (\gamma-1)-p}\right]^{\frac{3 \gamma}{3\gamma-2}}\!\right\}\!  , \quad \Theta(\rho, p) = \frac{p}{\tilde{k}\, n(\rho,p)}   . 
\label{s-idealgas}
\end{eqnarray}
\end{proposition}
The above generic ideal gas scheme corresponds to a specific choice of the functions $r(Q)$ and $s(Q)$ in proposition \ref{prop-scheme-parabolic-ideal}. Indeed, matching up expressions for $n$ and $s$ provided in (\ref{s-n-parabolic-ideal}) with those given in (\ref{n-idealgas}) and (\ref{s-idealgas}) we obtain:
\be \label{b-s-Q}
r(Q) = \tilde{e}_0 \, |Q|^{-\frac{1}{3 \gamma-2}} \, , \qquad s(Q) = \tilde{s}_0 + \frac{3 \tilde{k} \gamma}{3 \gamma-2} \ln |Q| \, .
\ee
Note that if we use (\ref{b-s-Q}) to determine the functions $\ell(Q)$ and $m(Q)$ given in (\ref{T-regular-b}), then the expression for $\Theta$ given in (\ref{T-regular}) is coherent with that given in (\ref{s-idealgas}).

Now we complete step 5' by analyzing for the generic ideal gas scheme the positivity conditions P and the compressibility condition H$_2$. As a consequence of (\ref{s-idealgas}), $\Theta>0$ when $n>0$. Moreover, $\rho > n>0$ if $e = \rho/n>1$, and (\ref{e-pi})  implies that this condition holds for a wide range of values of the arbitrary constant $e_0$ if $\pi < \gamma -1$.

On the other hand, in \cite{CFS-CC} we have shown that, for a generic ideal gas, the constraint H$_2$ can also be stated in terms of the hydrodynamic function of state $\chi=\chi(\pi)$: 
\be
\hspace{-5mm} {\rm H}^{\rm G}_2 : \qquad \qquad    \xi \equiv (2 \pi + 1) \chi(\pi) - \pi > 0 \, .
\ee
For the indicatrix function $\chi(\pi)$ given in (\ref{chi-parabolic-ideal}), we obtain:
\be
\xi(\pi) = \frac{\chi}{3 \gamma^2 \pi}[(6 \gamma^2 -3 \gamma -1) \pi^2 + (3 \gamma^2 -2 \gamma -2) \pi + (\gamma-1)] \, .
\ee
A straightforward calculation shows that $\xi(\pi)>0$ when $\pi < \gamma-1$. Consequently, if we take into account expression (\ref{ec-parabolic-ideal-0}) we can state:
\begin{proposition} \label{propo-cc2-ideal}
The ideal gas thermodynamic scheme associated with an ideal parabolic regular model fulfills the positivity condition {\em P} and the compressibility condition ${\rm H}_2$ if $(\gamma -1) \rho>p$, that is, in the spacetime domain where
\be \label{H2-parabolic-ideal}
\kappa^2 (9 \gamma^2 - 4) \, Q < -2 \phi^{3 \gamma-2}  .
\ee
\end{proposition}
%


\subsection{Models with the FLRW-limit temperature} 
\label{subsec-scheme-parabolic-ideal-FLRW}

The generic ideal gas thermodynamic scheme presented in the above section is just one of the possible thermodynamics that can be associated with each of the ideal parabolic regular solutions. As stated in proposition \ref{prop-scheme-parabolic-ideal}, these solutions model the evolution in l.t.e. of a wide family of perfect fluids defined by each choice of the two functions $s(Q)$ and $r(Q)$. 

As commented in subsection \ref{subsec-physical-non-perfect}, some of these choices lead to thermodynamic schemes that could also model the equilibrium state of inviscid fluids with a non-vanishing conductivity coefficient. Then, equation (\ref{Fourier}) applies and, for the geodesic flow of the SS solutions, it implies a homogeneous temperature, $\Theta = \Theta(t)$. Or, equivalently, $\Theta$ must be a function of $p$, a condition that is only consistent with the thermodynamic schemes that fulfill the conditions $\ell'(Q) = m'(Q)=0$.

Here we will restrict ourselves to the scheme with $\ell(Q) = 0$ and $m(Q) = m_0 \not = 0$, which leads to a model with the temperature of the FLRW limit, $\Theta \propto \phi^{-3(\gamma-1)}$ . With this choice, from (\ref{T-regular-b}) we obtain:
\be \label{rQsQ}
r(Q) = r_0 \, , \qquad s(Q) = s_0 + \frac{r_0}{m_0} Q  \, .
\ee
Then, if we take the arbitrary constants $r_0<0$ and $m_0>0$, we  obtain:
\begin{proposition} \label{propo-FLRW}
The ideal non-parabolic regular models admit thermodynamics with the same temperature as in the FLRW radiation model. The matter density $n$, the temperature $\Theta$ and the entropy $s$ are given by:
\begin{eqnarray}
n(\rho,p) = n_1 \, [\rho (\gamma-1)-p]\,  p^{\frac{1}{3 \gamma}}  , \qquad \Theta(\rho, p) = \Theta_1\, p^{1- \frac{1}{\gamma}}  \, , \label{n-FLRW} \\
s(\rho,p) = s_0 + s_1  \frac{\rho + (3 \gamma +1)p}{[\rho (\gamma-1)-p]\, p^{1- \frac{2}{3 \gamma}}}   , \quad s_1 \equiv  \frac{1}{(3 \gamma-2)n_1 \Theta_1} > 0   \, . 
\label{s-FLRW}
\end{eqnarray}
where $n_1$ and $\Theta_1$ are arbitrary positive constants.
\end{proposition}
It follows that $\Theta > 0$, and $n >0$ if $(\gamma -1) \rho>p$, that is, in the spacetime domain defined by constraint (\ref{H2-parabolic-ideal}). Moreover, for a wide range of the arbitrary positive constant $n_1$ we obtain $\rho > n$. Consequently, the positivity conditions P hold. On the other hand, expression (\ref{s-FLRW}) for $s$ is well defined, and from (\ref{pressure-parabolic-ideal}), expressions (\ref{n-FLRW}) can be written as:
\be \label{rhp-p-n}
\rho = \frac{1}{\gamma-1} (p + \frac{1}{n_1} n \,  p^{- \frac{1}{3 \gamma}}) \, , \qquad \Theta \propto \phi^{-3(\gamma-1)} \, ,
\ee
and we see that, indeed, we obtain the temperature of the $\gamma$-law models of the FLRW limit.

Finally we study the compressibility condition ${\rm H}_2$ for this thermodynamic scheme. From (\ref{s-FLRW}) we obtain $s'_{\rho}  < 0$, and thus, condition (\ref{H2-Theta}) holds. Consequently, we can state:
\begin{proposition} \label{propo-cc2-FLRW}
The thermodynamic scheme associated with an ideal parabolic regular model given in proposition \ref{propo-FLRW} fulfills the positivity conditions {\em P} and the compressibility condition ${\rm H}_2$ in the spacetime domain defined by the constraint (\ref{H2-parabolic-ideal}).
\end{proposition}
Note that the spatial domain where (\ref{H2-parabolic-ideal}) holds increases for early times for the expanding models, and it increases with time for contracting models.


\section{Ideal non-parabolic regular models}
\label{sec-ideal-non}

\subsection{Metric and hydrodynamic variables: energy density, pressure and speed of sound} 
\label{subsec-metric-ideal-non}

When $k=0$, case (iv) corresponds to case (ii) with $c_0 = 1$, $c_1 = 1/8$ and $c_3 = 3/8$. Then, the reasoning in subsection \ref{subsec-metric-parabolic-ideal} applies and the relation between the constants leads to a contradiction. Thus, the thermodynamic solutions will be non-parabolic models ($k^2 =1$). Then, equations (\ref{pressure-regular-tau}) and (\ref{p-dot}) for the functions $\phi(t)$ and $p(t)$ are equivalent to:
\be  \label{phi-punt-non}
p = \kappa^2  \phi^{-4}    \,   , \qquad \quad   \dot{\phi}^2  = \kappa^2 \, \phi^{-2} - k   \,  \, .
\ee
If we take into account these expressions, equations (\ref{xi-punt-1}) and (\ref{xi-punt-2}) for the functions $\beta(t)$ and $\xi(t)$ admit the sole solution:
\be \label{beta-0-non}
\beta = \beta(\phi) \equiv -k - \frac{\xi_0}{\phi} \, , \qquad   \dot{\xi} = - \frac{\xi_0 \,  \kappa^2}{\phi^3} \, , \qquad \xi_0 \not=0 \, .
 \ee
Note that the function $\beta$ appears in the metric expression (\ref{regular-SS-canonica}) through $\epsilon \beta + Q$. Thus we can redefine coordinate $z$ and functions $W(z)$,  $U(z)$, $V_1(z)$ and $V_2(z)$ from the old ones as $\xi_0 z$, $(W- k \epsilon/2)/\xi_0$, $(U + \epsilon/2)/\xi_0$, $V_1/\xi_0$ and $V_1/\xi_0$. Then, $U=  - k W$, and we can take $\beta = - \phi^{-1}$, $\dot{\xi} = - \frac{\kappa^2}{\phi^3}$.  Moreover, (\ref{phi-punt-non}) is the Friedmann equation for a non-parabolic radiation FLRW-model, which can easily be integrated. Then, considering (\ref{density-regular}) and (\ref{beta-0-non}), we obtain:
\begin{proposition} \label{prop-regular-nonparabolic-ideal}
The ideal non-parabolic regular models have a metric line element of the form (\ref {regular-SS-canonica}), where $Q = SC$, $C$ and $S$ taking the form (\ref{II-C}) and (\ref{II-S}), with $k=\pm 1$ and $U=  - k W$, and where  $\phi(t)$ and $\beta(t)$ are given, respectively, by:
\begin{equation} \label{phi-t-non}
\phi(t) = \left[k(\kappa^2 - (t-t_0)^2) \right]^{\frac{1}{2}}   ,  \qquad  \beta = \beta(\phi) \equiv - \frac{1}{\phi} \,  .
\end{equation}
Moreover the pressure $p$ and the energy density $\rho$ are:
\begin{eqnarray} \label{pressure-non-parabolic-ideal}
p = \frac{ \kappa^2}{\phi^{4}}    \,  ,  \\[1mm]
\rho = \frac{3 \kappa^2 }{\phi^{4}} 
\left[1 + \frac{2 \varepsilon }{3 (\phi  Q - \varepsilon)}\right]  \, .
\label{density-non-parabolic-ideal}
\end{eqnarray}
And the speed of sound is given by: 
\be \label{chi-non-parabolic-ideal}
c_s^2  = \chi(\pi)  \equiv \frac{8 \, \pi^2}{ (\pi + 1) (3  \pi  + 1)} \, , \qquad \pi \equiv \frac{p}{\rho} \, .
\ee
\end{proposition}
%


\subsection{Analysis of the solutions. Energy conditions} 
\label{subsec-energy-c-non}

It is worth remarking the following qualities of these solutions:
\begin{itemize}
\item[(i)] 
The metric depends on three arbitrary functions of $z$, $V_1 (z)$, $V_2 (z)$ and $W(z)$, and an effective parameter, $\kappa$, whose square determines the strength of the density $\rho$ and the pressure $p$. The constant $t_0$ only determines an origin of time and it does not affect the metric.
\item[(ii)] 
For $k=-1$, expanding models evolve to the open FLRW radiation models.
\end{itemize}

Now we achieve the analysis of the energy conditions E$^{\rm G}$ by obtaining the spacetime domains these conditions (\ref{e-c-G}) are fulfilled. From expressions (\ref{pressure-non-parabolic-ideal}) and (\ref{density-non-parabolic-ideal}) for the pressure and density we obtain ($\varepsilon = 1$):
\be \label{ro/p}
\pi^{-1} = \frac{\rho}{p} =  3 + \frac{2}{Q \phi -1} \,  \, .
\ee
Consequently, $\rho > p$ if either $Q \phi -1 > 0$ or $Q < 0$, so that:
\begin{proposition} \label{prop-ec-non-parabolic-ideal}
The ideal non-parabolic regular models in proposition \ref{prop-regular-nonparabolic-ideal} fulfill the energy conditions {\rm E}$^{\rm G}$ given in (\ref{e-c-G}) in the spacetime domains where one of the following two conditions holds:
\be \label{ec-non-parabolic-ideal}
Q \phi -1 > 0 \, , \qquad  Q < 0  \, .
\ee
\end{proposition} 
Note that, for an expanding epoch, the spatial domain where conditions (\ref{ec-non-parabolic-ideal}) hold increases with time.

%


\subsection{Compressibility conditions} 
\label{subsec-compress-c-non-parabolic-ideal}

Now we will complete step 3' for the ideal non-parabolic regular models by analyzing the compressibility conditions H$_1^{\rm G}$.  From the expression (\ref{chi-non-parabolic-ideal}), a straightforward calculation leads to:
\be \label{chi-prima-non}
\chi' (\pi) = \frac{16 \pi (2 \pi + 1)}{[(\pi + 1) (3  \pi  + 1)]^2} > 0  \, ,  \qquad \chi(0) =0 \, , \quad \chi(1) = 1 \, .
\ee
Thus, the first compressibility condition H$_1^{\rm G}$ in (\ref{cc-ideal}) holds for any $\pi \in ]0,  1[$. And then, we have $\zeta > 0$, and the second compressibility condition H$_1^{\rm G}$ holds. So, we have shown:
\begin{proposition} \label{prop-cc-non-parabolic-ideal}
The ideal non-parabolic regular models in proposition \ref{prop-regular-nonparabolic-ideal} fulfill the compressibility conditions ${\rm H}_1^{\rm G}$  in the domain where the energy conditions, $\pi \in ]0,  1[$, hold. 
\end{proposition} 
%


\subsection{Thermodynamic schemes: entropy, matter density and temperature} 
\label{subsec-scheme-non-parabolic-ideal}

Now we study the full set of thermodynamics associated with the (strict, $\varepsilon =1$) ideal non-parabolic singular models. We must particularize the thermodynamic schemes presented in subsection \ref{subsec-scheme-regular} for the solutions in proposition \ref{prop-regular-nonparabolic-ideal}. Note that, from (\ref{phi-t-non}), (\ref{pressure-non-parabolic-ideal}) and (\ref{ro/p}) we obtain :
\be \label{Q-rho-p-1-non}
Q = Q(\rho,p) \equiv   \frac{\rho -p}{\sqrt{\kappa}(\rho - 3 p)} p^{1/4} \, , \qquad  \phi^3 ( \beta + Q) =  \frac{ 2 \kappa \sqrt{p}}{\rho - 3 p}  \, .
\ee

On the other hand, the functions $\lambda(t)$ and $\mu(t)$ defined in (\ref{lambda-mu}) can be computed in terms of $\phi$ and in terms of $p$:
\begin{equation} \label{lambda-mu-p-non}
\lambda =  -\frac{2 \kappa^2}{\phi^2} = -2 \kappa  \, \sqrt{p} \,  ,  \qquad   \mu =  \frac{4 \kappa^2}{\phi} = 4 \kappa^{3/2}  \, p^{1/4}  \, . 
\end{equation}
Then, substituting all these formulas in expressions (\ref{s-n-regular}), (\ref{T-regular}) and (\ref{T-regular-b}) for $n$, $s$ and $\Theta$, we obtain:
\begin{proposition} \label{prop-scheme-non-parabolic-ideal}
The thermodynamics associated with the ideal parabolic regular models given in proposition \ref{prop-regular-parabolic-ideal} are determined by a specific entropy $s$ and a matter density $n$ of the form:
\be  \label{s-n-non-parabolic-ideal}
 s = s(Q) \equiv s(\rho, p) \, , \qquad  n =  \frac{\rho - 3 p}{2 \kappa \, r(Q) \sqrt{p}} \equiv n(\rho,p)  \, ,  
\ee
where $s(Q)$, $s'(Q)\not=0$, and $r(Q)\not=0$ are arbitrary real functions of the function of state $Q=Q(\rho,p)$ given in (\ref{Q-rho-p-1-non}). Moreover the temperature is of the form (\ref{T-regular}), where $\lambda(p)$ and $ \mu(p)$ are given in (\ref{lambda-mu-p-non}), and  $\ell(Q)$ and $m(Q)$ are given in (\ref{T-regular-b}).
\end{proposition}


\subsection{Models with a generic ideal gas thermodynamic scheme} 
\label{subsec-scheme-parabolic-ideal-ideal}

Now we analyze the thermodynamic scheme modeling a generic ideal gas that can be associated with ideal non-parabolic regular models. From the general expressions obtained in lemma 4 of \cite{CFS-LTE}, we establish that the {\em specific energy density} is given by:
\be 
e = \frac{\rho}{n} = e(\pi) \equiv e_0 \frac{1-3 \pi}{(\pi-1)^2}\,  ,   \label{e-pi-non}
\ee
and the other thermodynamic quantities are given by:
\begin{proposition} \label{esquema-ideal-non}
The matter density $n$, the specific entropy $s$ and the temperature $\Theta$ of the generic ideal gas scheme associated with an ideal non-parabolic regular model take the expressions: 
\begin{equation}
\hspace{-22mm} n(\rho,p) \! = \! \frac{(\rho-p)^2}{e_0(\rho - 3p)}    , \quad 
s(\rho,p)\!  = \! s_0 \! + \! \tilde{k} \ln \!  \left\{\! \frac{1}{p} \left[\frac{\rho - 3 p}{p - \rho}\right]^{4}\!\right\}\!  , \quad \Theta(\rho, p) \! = \! \frac{p}{\tilde{k}\, n(\rho,p)}   . 
\label{n-s-idealgas-non}
\end{equation}
\end{proposition}
The above generic ideal gas scheme corresponds to the following specific choice of the functions $r(Q)$ and $s(Q)$ in proposition \ref{prop-scheme-parabolic-ideal}:
\be \label{b-s-Q-non}
r(Q) = \tilde{e}_0 \, Q^{-2} \, , \qquad s(Q) = \tilde{s}_0 - 4 \tilde{k} \ln |Q| \, .
\ee

Now we complete step 5' by analyzing the positivity conditions P and the compressibility condition H$_2$ for the generic ideal gas scheme. As a consequence of (\ref{n-s-idealgas-non}), $\Theta>0$ when $n>0$. Moreover, $\rho > n>0$ if $e = \rho/n>1$, and (\ref{e-pi-non})  implies that this condition holds for a wide range of values of the arbitrary constant $e_0$ if $\pi < \tilde{\pi}_{M} < 1/3$.

On the other hand, we must impose the compressibility condition for a generic ideal gas {\rm H}$^{\rm G}_2$. For the indicatrix function $\chi(\pi)$ given in (\ref{chi-non-parabolic-ideal}), we obtain:
\be \label{xi-non}
\xi(\pi) = \frac{\chi}{8 \pi}[13 \pi^2 + 4 \pi -1] > 0 \, , \quad {\rm if} \quad \pi > \tilde{\pi}_m = \frac{1}{13}(\sqrt{17}-2) \, .
\ee
Consequently, if we take into account expression (\ref{ro/p}) we can state:
\begin{proposition} \label{propo-cc2-ideal-non}
The ideal gas thermodynamic scheme associated with an ideal non-parabolic regular model fulfills the positivity conditions {\em P} and the compressibility condition ${\rm H}_2$ if $p/\rho \in ]\tilde{\pi}_m, \tilde{\pi}_M[$, that is, in the spacetime domain where
\be \label{H2-non-parabolic-ideal}
 q_M < Q \,  \phi <  q_m \, , \quad q_M \equiv 1 + 2/(\tilde{\pi}_M - 3)\, , \quad q_m \equiv 1 + 2/(\tilde{\pi}_m - 3).
\ee
\end{proposition}
%


\subsection{Models with the FLRW-limit temperature} 
\label{subsec-scheme-parabolic-ideal-FLRW}

We can also interpret the ideal non-parabolic regular solutions as a thermodynamic scheme compatible with a non-vanishing conductivity coefficient, and with the homogeneous temperature of the FLRW limit (the radiation model), $\Theta \propto \phi^{-1}$. We must again choose $\ell(Q) = 0$ and $m(Q) = m_0 \not = 0$ in expression (\ref{T-regular}). With this choice, from (\ref{T-regular-b}) we obtain the expressions (\ref{rQsQ}) for the functions $r(Q)$ and $s(Q)$. Then, if we take the arbitrary constants $r_0>0$ and $m_0>0$, we  obtain:
\begin{proposition} \label{propo-FLRW-non}
The ideal non-parabolic regular models admit thermodynamics with the same temperature as in the radiation model of the FLRW limit. The matter density $n$, the temperature $\Theta$ and the entropy $s$ are given by:
\begin{eqnarray}
n(\rho,p) = n_1 \, \frac{\rho - 3 p}{ \sqrt{p}}  , \qquad \Theta(\rho, p) = \Theta_1\, p^{1/4}  \, , \label{n-FLRW-non} \\
s(\rho,p) = s_0 + s_1  \frac{\rho -p}{\rho - 3 p} p^{1/4}  , \quad s_1 \equiv  \frac{2}{n_1 \Theta_1} > 0   \, . 
\label{s-FLRW-non}
\end{eqnarray}
where $n_1$ and $\Theta_1$ are arbitrary positive constants.
\end{proposition}
It follows that $\Theta > 0$, and $n >0$ if $ \rho>3p$. Moreover, for a wide range of the arbitrary positive constant $n_1$ we obtain $\rho > n$. Consequently, the positivity conditions P hold. On the other hand, expression (\ref{s-FLRW}) for $s$ is well defined, and from (\ref{pressure-parabolic-ideal}), expressions (\ref{n-FLRW}) can be written as:
\be \label{rhp-p-n-non}
\rho =  3 p + \frac{1}{n_1} n \,  \sqrt{p} \, , \qquad \Theta \propto \phi^{-1} \, ,
\ee
and we see that, indeed, we obtain the temperature of the radiation FLRW models.

Finally we study the compressibility condition ${\rm H}_2$ for this thermodynamic scheme. From (\ref{s-FLRW-non}) we obtain $s'_{\rho}  < 0$, and thus, condition (\ref{H2-Theta}) holds. Consequently, we can state:
\begin{proposition} \label{propo-cc2-FLRW-non}
The thermodynamic scheme associated with an ideal non-parabolic regular model given in proposition \ref{propo-FLRW-non} fulfills the positivity conditions {\em P} and the compressibility condition ${\rm H}_2$ if  $\rho>3p$, that is, in the spacetime domain defined by the constraint $Q \phi > 1$.
\end{proposition}
Note that this spatial domain increases at early epoch for contracting models, whereas it increases with time for expanding models.


\section{Summary and work in progress}
\label{sec-conclusions}

In this paper we have accomplished the task started in \cite{PSS}: the study of class II Szekeres-Szafron metrics without symmetries, which can be interpreted as a thermodynamic perfect fluid in local thermal  equilibrium. Our analysis leads to two families, the singular models (considered in \cite{PSS}) and the regular models (considered here). The thermodynamic schemes associated with these thermodynamic solutions have been outlined. We have analyzed in depth the solutions with a significant physical interpretation: those compatible with the equation of state of a generic ideal gas. In the case of the regular models two subfamilies must be distinguished, the {\em ideal parabolic regular models} ($k=0$) and the {\em ideal non-parabolic regular models} ($k=\pm 1$). Here we have analyzed the macroscopic constraints for physical reality (energy conditions, positivity conditions and compressibility conditions) for these two subfamilies, and a similar analysis has been carried out in \cite{PSS} for the {\em ideal singular models}.

In local thermal equilibrium, the indicatrix function $\chi$ is a function of state  which gives the square of the speed of sound in terms of the hydrodynamic quantities $\rho$ and $p$, $c_s^2 = \chi(\rho, p)$. This function plays an important role in our study of the thermodynamic properties of the solutions. In dealing with ideal models the indicatrix function depends on the sole hydrodynamic quantity $\pi=p/\rho$, $\chi= \chi(\pi)$, and on an adiabatic-like index $\gamma \in ]1,2[$. The ideal non-parabolic regular models are only compatible with $\gamma = 4/3$ (ultra-relativistic gas), and the indicatrix function is given by the function $\chi_1(\pi)$ given in (\ref{chi-non-parabolic-ideal}). It is worth considering the two other families of ideal models for $\gamma = 4/3$. In this case, the singular models have precisely the same expression $\chi_1(\pi)$ for the indicatrix function. We have $\chi_1(0)=0$ and $\chi_1(1/3)=1/3$, which are the values that take the indicatrix function, $\chi_s(\pi)$, of a relativistic Synge gas. But, $\chi_1'(0) \not= \chi_s'(0)$ and $\chi_1'(1/3)  \not= \chi_s'(1/3)$. Thus, the ideal singular models (with $\gamma = 4/3$) and the ideal non-parabolic regular models approach the Synge gas at low and high temperatures, but only at zero order (see Figure 1).

On the other hand, the parabolic regular models with $\gamma = 4/3$ have an indicatrix function $\chi_2(\pi)$ that can be obtained from (\ref{chi-parabolic-ideal}). This function does not fulfill the compressibility conditions for small values of $\pi$, but $\chi_2(1/3)  = \chi_s(1/3)=1/3$ and $\chi_2'(1/3)  = \chi_s'(1/3)=1/2$. Thus, the ideal parabolic regular models (with $\gamma = 4/3$) are not valid for describe low temperature states, but they approach at first order the Synge gas at high temperatures (see Figure 1).

\begin{figure}[t]
\centerline{
\includegraphics[width=0.70\textwidth]{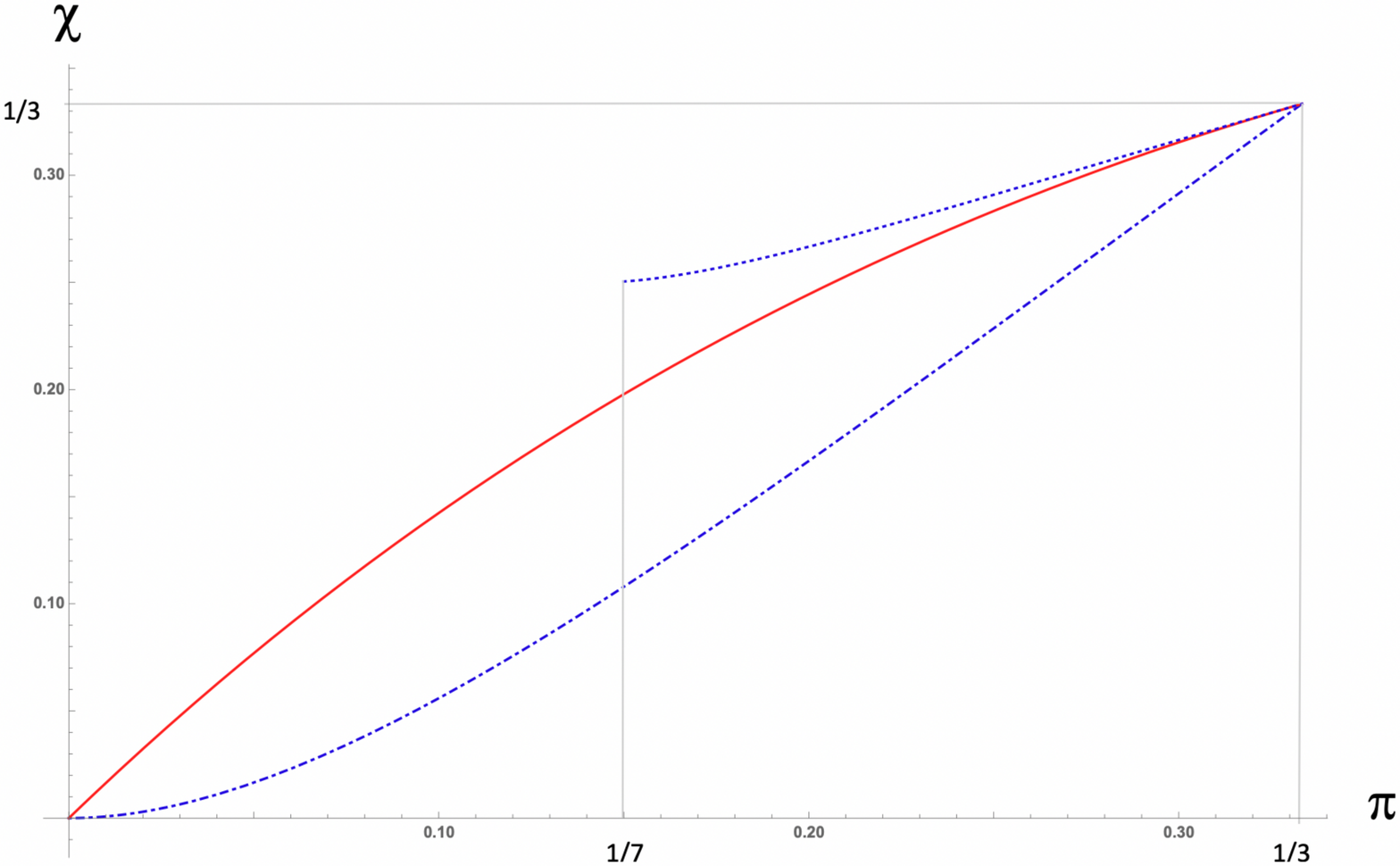}}
\caption{Graph of the function $\chi= \chi(\pi)$ that gives the square of the speed of sound $c_s^2 = \chi$ as a function of the hydrodynamic quantity $\pi=p/\rho$ for the ideal models with $\gamma = 4/3$ in the interval $[0,1/3]$. The red solid line shows this function for the relativistic Synge gas. The blue dotdashed line shows $c_s^2 = \chi_1(\pi)$, the square of the speed of sound for both the ideal singular models and the ideal non-parabolic regular models. The blue dotted line shows $c_s^2 = \chi_2(\pi)$, the square of the speed of sound for the ideal parabolic regular models, in the interval $]1/7, 1/3[$ where it is an increasing function. Note that $\chi_1(\pi)$ approaches at zero order the Synge gas at low and at high temperatures. And $\chi_2(\pi)$ approaches at first order the Synge gas at high temperatures.
\label{Fig-1}}
\end{figure}

It is worth remarking that both the non-parabolic regular models and the singular models evolve to the (homogeneous) FLRW $\gamma$-law models in the expanding case, and the inhomogeneities increase with time in the contracting models. Nevertheless, the
parabolic regular models evolve to the (homogeneous) FLRW $\gamma$-law models in the contracting case, and the inhomogeneities increase with time in the expanding models.

Further work will be devoted to analyzing the Szekeres-Szafron solutions of class II with a G$_3$ group of isometries. They are necessarily thermodynamic solutions, but two basic problems remain open: (i) the determination of the associated thermodynamic schemes, and (ii) the detailed analysis of the ideal gas models. Moreover, these studies will allow us to interpret the (barotropic) Kantowski-Sachs metrics as a thermodynamic perfect fluid in isentropic evolution.


\ack This work has been partially supported by the Spanish ``Ministerio de
Econom\'{\i}a y Competitividad", MINECO-FEDER project FIS2015-64552-P.

\appendix

\section{Proof of lemma \ref{lemma-equations}}
\label{Ap-equations}

We are going to study here the differential system (\ref{p-dot}), (\ref{xi-punt-1}), (\ref{xi-punt-2}) for the two functions $\phi(t), \beta(t)$. From (\ref{pressure-regular-tau}) we obtain:
\begin{eqnarray}
\left(\frac{\dot{\phi}}{\phi}\right)^{\! \! .} = - \frac12 (p+\tau) - q \, , \qquad q \equiv \frac{\dot{\phi}^2}{\phi^2} \, ,  \label{q} \\
\dot{\tau} = - \frac{\dot{\phi}}{\phi}(p + 3 \tau) \, , \qquad  \dot{q} = - \frac{\dot{\phi}}{\phi}(p + \tau + 2 q)  \, . \label{q-punt}
\end{eqnarray}
We can eliminate $\beta$ from (\ref{xi-punt-1}) and (\ref{xi-punt-2}) if we differentiate these equations and make use of (\ref{q}) and (\ref{q-punt}). Then, we obtain the following 4th-order differential equation for the sole function $\phi$:
\be \label{p-ddot-1}
\ddot{p} = \frac{1}{2 c_1} p^2 + \frac{\dot{\phi}}{\phi} \, \dot{p} - [(c_1\! - \! 2)p +3 c_1 \tau] \, \frac{\dot{p}^2}{p^2} \,  . 
\ee
Now we analyze the compatibility of the equations (\ref{p-dot})  and (\ref{p-ddot-1}) for $\phi$. If we differentiate (\ref{p-dot}) and take into account (\ref{q}) and (\ref{q-punt}) we obtain:
\be \label{p-ddot-2}
(c_3 p + 3 c_1 \tau) \ddot{p} = \frac{3}{2} (p+ \tau + 2 q)\, p^2 + 3 \frac{\dot{\phi}}{\phi} [(c_1\!  - \! 2)p +3 c_1 \tau] \, \dot{p} - c_3 \, \dot{p}^2 \,  . 
\ee
Then, from (\ref{p-ddot-1}), (\ref{p-ddot-2}) and (\ref{p-dot}) we can eliminate $\ddot{p}$ and obtain:
\be \label{p-dot-2}
c_3 \, \dot{p}^2 =  6 \, p^2 (c_4 \,  p +  q)  \,  ,   \qquad c_4 \equiv \frac14\! \left(\! 1 \!- \! \frac{c_3}{3c_1} \! \right) \, . 
\ee
And from (\ref{p-dot}) and (\ref{p-dot-2}) we obtain:
\begin{equation} \label{p-pol3-1}
a_3 \, p^3 + a_2 \, p^2 + a_1 \, p + a_0 =0  \,  , 
\ee
\be \label{p-pol3-1b}
\hspace{-1cm}\begin{array}{lllll}
a_3 \equiv  2 c_3^2 c_4   \, , & a_2 \equiv  c_3 [2 c_4 \, \tau + (2 c_3 - 3) \, q] \, ,   \\
a_1 \equiv   6 c_1\, \tau \,   [3 c_1 c_4\, \tau + 2 c_3  \, q]    \, , \qquad & a_0 \equiv 18 c_1^2 \, \tau^2 \, q  \, .   \\
\end{array}
\end{equation}
Moreover, by differentiating (\ref{p-dot-2}) and by using (\ref{p-dot}) it follows:
\be \label{p-ddot-3}
c_3 \, \ddot{p} = (9 c_4 \!+ \!c_3)\, p^2 +  [(c_3 \!+ \! 3c_1)\tau  + 2(3 \!+ \! c_3) q] \, p + 3 c_1 \tau (\tau + 2q) \,  . 
\ee
Then, from this equation and (\ref{p-dot}), (\ref{p-ddot-1}) and (\ref{p-dot-2}) we obtain:
\begin{equation} \label{p-pol3-2}
b_3 \, p^3 + b_2 \, p^2 + b_1 \, p + b_0 =0  \,  , 
\ee
\be \label{p-pol3-2b}
\hspace{-1.6cm}
\begin{array}{ll}
b_3 \equiv  \frac14 c_3 [3(2c_1\!-\!1)\!+\!2c_3 \!-\!  c_3/c_1] \,  , \qquad \quad  \ \,  & b_2 \equiv m_2 \, \tau  + n_2 \, q  \, ,  \\
b_1 \equiv 3 c_1 \, \tau[m_1 \, \tau  + n_1 \, q] \,  ,  & b_0 \equiv 9 c_1^2 \, \tau^2 \, (\tau+8q)   \, . \quad 
\end{array}
\end{equation}
\be \label{p-pol3-2c}
\hspace{-1.6cm}
\begin{array}{ll}
m_2 \equiv \frac14[ 9c_1(4c_3 \! + \! 2c_1 \!   - \!  1) \!  - \! c_3(2c_3 \! + \! 3)]  \, , \quad & n_2 \equiv c_3(6c_1 \!   + \!   2c_3 \! - \!   3) \, ,  \\
 m_1 \equiv \frac12 (15 c_1 +c_3)  \, , & n_1 \equiv 6(c_1\! -\!1) \!+\! 10 c_3  \, .  
\end{array}
\end{equation}

From the two third-degree polynomial equations in $p$ (\ref{p-pol3-1}) and (\ref{p-pol3-2}) we can obtain the following two second-degree polynomial equations:
\begin{equation} \label{p-pol2}
e_2 \, p^2 + e_1 \, p + e_0 =0  \,  , \qquad \quad 
d_2 \, p^2 + d_1 \, p + d_0 =0  \,  , 
\end{equation}
\be \label{p-pol2-a}
\hspace{-1.6cm}
\begin{array}{lll}
e_2 \equiv a_2 b_3 -a_3 b_2 \,  , \quad \quad    & e_1 \equiv a_1 b_3 -a_3 b_1 \,  , \quad \quad    & e_0\equiv a_3 b_0 - a_0 b_3 \,    \\
d_2 \equiv c_0\,  , \quad \quad   & d_1 \equiv a_2 b_0 -a_0 b_2  \,  , \quad \quad    & d_0 \equiv a_1 b_0 -a_0 b_1   \, .  
\end{array}
\ee
The two polynomial equations (\ref{p-pol2}) admit a common root when:
\be  \label{compat-2}
(e_0 d_2 - e_2 d_0)^2 = (d_2e_1-e_0 d_1)(e_0 d_1 -e_1 d_0) \, .
\ee
Taking into account definitions (\ref{p-pol3-1b}), (\ref{p-pol3-2b}) and (\ref{p-pol3-2c}), condition (\ref{compat-2}) states the vanishing of a fifth-degree polynomial for the quotient $\tau/q$ with coefficients depending on the constants $c_1$ and $c_3$:
\be  \label{pol-tau/q}
\sum_{n=0}^5 f_n (\tau/q)^n =0 \, , \qquad  f_n \equiv f_n(c_1, c_3) \, .
\ee

When $\tau = c_5 \, q$, $c_5 = constant$, equation (\ref{pol-tau/q}) is a constraint for the constants $c_1$, $c_3$ and $c_5$. Moreover, from the expressions (\ref{pressure-regular-tau}) and (\ref{q}) we have that either $\ddot{\phi}=0$ (case (i)) or $k=0$ and $\tau=q$. In this last case, (\ref{p-pol3-1}) and (\ref{p-pol3-2}) become polynomial equations for the quotient $\tau/p$. The coefficients, depending on $c_1$ and $c_3$, cannot vanish identically and, consequently, $k=0$ and $\tau = c_0 \, p$ (case (ii)). 

Otherwise, when $\tau \not= c_5 \, q$, a necessary condition for (\ref{pol-tau/q}) to be met is that all the coefficients $f_n$ vanish: 
\be \label{fn}
f_n(c_1, c_3)=0 \, .
\ee
This system of six equations for $c_1$ and $c_3$ admits two solutions: ($c_1= 2, c_3=6$) and ($c_1= 1/8, c_3=3/8$). If we substitute these values in the polynomial equations (\ref{p-pol3-1}) and (\ref{p-pol3-2}) we obtain, respectively, $p=-2\tau$ and $p= \tau$ (cases (iii) and (iv)). 


\section{Analyses of the cases (i) and (iii)}
\label{Ap-rarets}

Case (i): $\ddot{\phi}=0$. Then, we have $\dot{\phi}=f = constant$, and from (\ref{pressure-regular-tau}) we obtain $p= - \tau = -(f^2+k)\phi^{-2}$. This expression is compatible with a positive pressure only when $k=-1$, and then:
\be \label{zeta}
p = \frac{A_1^2}{\phi^2} \, , \qquad \quad  A_1^2 \equiv 1 - f^2 > 0 \, .
\ee
This expression is only compatible with equation (\ref{p-dot}) when $c_3 = 3(c_1 + \frac12)$. On the other hand, equation (\ref{xi-punt-2}) implies
\be
\xi = A_2 \phi^{\nu} \, , \qquad  \quad \nu \equiv - \frac{A_1^2}{4 f^2 c_1} \, .
\ee
Then, the first equation in (\ref{xi-punt-1}) leads to $\beta= 1- f A_2(\nu+1) \phi^{\nu-1}$, and the second equation in (\ref{xi-punt-1}) implies:
\be
1+ f^2(\nu^2-1) = 0 \, ,
\ee
which is not compatible with (\ref{zeta}).\\[3mm]
Case (iii): $c_1= 2, \ c_3=6$ and $p=-2\tau$. If we take into account (\ref{pressure-regular-tau}) and consider a positive pressure, equation (\ref{p-dot}) implies $k=-1$ and
\be \label{zeta-hat}
p = \frac{\hat{\varsigma}^2}{\phi} \, , \qquad \quad  \dot{\phi}^2  = 1 -\frac{\hat{\varsigma}^2}{2} \phi \, , 
\ee
where $\hat{\varsigma}$ in an arbitrary non-vanishing constant. Then, we can integrate equation (\ref{xi-punt-2}) for $\xi$ and obtain:
\be \label{zeta-tilde}
\xi^2 = \frac{A_3^2}{1 -\frac{\hat{\varsigma}^2}{2} \phi} \, , 
\ee
$A_3$ being an arbitrary constant. If we study the compatibility of (\ref{zeta-hat}) and (\ref{zeta-tilde}) with equation (\ref{xi-punt-1}) we obtain $\beta=1$ and $\xi =0$, and then $\rho = 3 \tau = - \frac32 p < 0$.


\end{document}